\documentclass[12pt]{article}
\usepackage{amsmath}
\usepackage{amsfonts}
\usepackage{graphicx,psfrag,epsf}
\usepackage{enumerate}
\usepackage[backend=bibtex,style=nature]{biblatex}
\usepackage{url} 
\usepackage{authblk}

\usepackage{mdframed}
\usepackage{caption}
\usepackage{subcaption}
\usepackage{siunitx}
\usepackage{float}
\usepackage[normalem]{ulem}
\newcommand{\stkout}[1]{\ifmmode\text{\sout{\ensuremath{#1}}}\else\sout{#1}\fi}
\usepackage{mathtools}
\usepackage{mdwlist}
{
\begin{basedescript}{
\desclabelwidth{2em}}}
{
\end{basedescript}
}

\usepackage[dvipsnames]{xcolor} %
\usepackage{diagbox}
\usepackage{bbm}
\usepackage{algorithm}
\usepackage[noend]{algpseudocode}
\algrenewcommand\algorithmicrequire{\textbf{Input:}}

\algrenewcommand\algorithmicensure{\textbf{Output:}}

\usepackage{amsthm}
\usepackage{mathtools}

\usepackage{hyperref}
\hypersetup{
    colorlinks=true,
    linkcolor=blue,
    citecolor=blue,
    filecolor=magenta,      
    urlcolor=blue,
}

\makeatletter
\def\BState{\State\hskip-\ALG@thistlm}

\newcommand\bmu{\mbox{\boldmath${\mu}$}}

\newcommand\btheta{\mbox{\boldmath${\theta}$}}

\newcommand\ba{{\bf a}}

\newcommand\bs{{\boldsymbol s}}

\newcommand\bV{{\bf V}}

\newcommand\bW{{\bf W}}

\newcommand\bX{{\bf X}}
\newcommand\bx{{\bf x}}
\newcommand\bY{{\bf Y}}

\newcommand\bZ{{\bf Z}}
\newcommand\bz{{\bf z}}

\newcommand{\bOmega}{\boldsymbol{\Omega}}
\newcommand{\bomega}{\boldsymbol{\omega}}
\newcommand{\bphi}{\boldsymbol{\phi}}

\newcommand{\bzeta}{\boldsymbol{\zeta}}

\newcommand{\sumK}{\sum_{k=1}^K}
\newcommand{\prodK}{\prod_{k=1}^K}
\newcommand{\sumN}{\sum_{j=1}^{n_s}}

\makeatletter
\renewcommand*\env@matrix[1][\arraystretch]{%
  \edef\arraystretch{#1}%
  \hskip -\arraycolsep
  \let\@ifnextchar\new@ifnextchar
  \array{*\c@MaxMatrixCols c}}
\makeatother

\begin{document}
\title{\bf Capturing Extreme Events in Turbulence using an Extreme Variational Autoencoder (xVAE)}
\date{}
\author[1,*]{Liken Zhang}
\author[2]{Kiran Bhaganagar}
\author[1]{Christopher K. Wikle}
\affil[1]{Department of Statistics, University of Missouri}
\affil[2]{Laboratory of Turbulence, Sensing and Intelligence Systems, Department of Mechanical Engineering, The University of Texas at San Antonio (UTSA)}
\affil[*]{\textit{Corresponding author}: Likun Zhang, likun.zhang@missouri.edu}

\maketitle
\begin{abstract}
   Turbulent flow fields are characterized by extreme events that are statistically intermittent and carry a significant amount of energy and physical importance. To emulate these flows, we introduce the extreme variational Autoencoder (xVAE), which embeds a max-infinitely divisible process with heavy-tailed distributions into a standard VAE framework, enabling accurate modeling of extreme events. xVAEs are neural network models that reduce system dimensionality by learning non-linear latent representations of data. We demonstrate the effectiveness of xVAE in large-eddy simulation data of wildland fire plumes, where intense heat release and complex plume-atmosphere interactions generate extreme turbulence. Comparisons with the commonly used Proper Orthogonal Decomposition (POD) modes show that xVAE is more robust in capturing extreme values and provides a powerful uncertainty quantification framework using variational Bayes. Additionally, xVAE enables analysis of the so-called copulas of fields to assess risks associated with rare events while rigorously accounting for uncertainty, such as simultaneous exceedances of high thresholds across multiple locations. The proposed approach provides a new direction for studying realistic turbulent flows, such as high-speed aerodynamics, space propulsion, and atmospheric and oceanic systems that are characterized by extreme events.
   
\end{abstract}
\noindent%

\section{Introduction}

\par Turbulence is characterized by chaotic, unpredictable fluid motion \cite{kolmogorov1991local,frisch1995turbulence}. With increasing Reynolds numbers, the extreme events---high intensity, rare occurrences with the formation of strong localized structures---become more dominant \cite{yeung2015extreme,moffatt2021extreme}. At high Reynolds numbers, extreme events such as large vortex formations, intense vorticity, intermittent bursts of energy strongly influence the behavior of turbulent flows \cite{goldenfeld2010extreme,buaria2019extreme}. The fluctuations in velocity, vorticity, pressure, and other turbulent quantities in these flows are often described by statistical distributions. Near the time-averaged mean (in the core of the distribution), turbulent fluctuations often exhibit approximately Gaussian behavior. However, as fluctuations grow, their statistical properties deviate significantly from the bulk \cite{l2001outliers}, and this becomes more pronounced with increasing Reynolds number. 


\par Accurate characterization of extreme fluctuations (i.e., the tail behavior of turbulence distributions) is critical due to their impact on risk assessment and system design \cite{sapsis2021statistics,gotoh2022transition}. In atmospheric sciences, the tail of the turbulence distribution influences the prediction of extreme weather events such as hurricanes or typhoons, where localized regions of extreme turbulence can cause widespread damage \cite{fukami2023grasping,sapsis2021statistics}. 
In aerospace engineering, extreme turbulence events (e.g., clear air turbulence) pose significant risks to aircraft \cite{sharman2016nature}. Understanding the tail of the turbulence distribution helps engineers design aircraft that can withstand these rare but dangerous conditions. In high-speed flows with combustion \cite{driscoll2020premixed}, the shock waves and detonation waves are the extreme events that occur during a short time period and within a narrow region and are dominant in the tail region of turbulence \cite{hassanaly2021classification}. Capturing extreme detonation and shock events is critical for supersonic and hypersonic aerodynamic and rocket propulsion systems. 

\par The last 40 years have witnessed significant efforts by the turbulence community to develop statistical methods that emulate turbulent flow fields and capture the essential features of the complete system with far fewer degrees of freedom at a fraction of the computational cost \cite{farazmand2017variational,duraisamy2019turbulence,zare2020stochastic, beck2021perspective}. 
Proper Orthogonal Decomposition (POD) has been a successful statistical method for dimensional reduction that extracts dominant energetic structures in the turbulent flow field as the eigen-functions of the two-point correlation tensor of the velocity fields  \cite{Lumley93}. Once the basis has been determined, the turbulent flow fields are decomposed into modes represented by    
the coefficients of the expansion basis. 
Only the most energetic POD modes are retained in the projection and modes with negligible energy content are discarded by truncating the 
expansion. Recently, variations of POD such as spectral POD \cite{cammilleri2013pod,hijazi2020data, schubert2022towards,schmid2022dynamic}, dynamical model decomposition, and various others have gained popularity for reducing the dimensionality.  However, because POD and its variants rely on linear subspaces, they excel at approximating flows that can be well-represented by linear components but struggle with capturing intermittent and extreme turbulence. In particular, buoyant plumes often exhibit large, anisotropic clusters of extremes, which tend to be dampened when only a limited number of linear modes is retained.

More recently, machine learning techniques are gaining popularity for developing reduced-order models  \cite{baddoo2023physics,fukami2020convolutional,momenifar2022physics,glaws2020deep}. Of these architectures,  variational autoencoders \cite[VAEs;][]{kingma2013auto} have proven to be effective for encoding the spatial information of fluid flows in non-linear low-dimensional latent-space \cite{solera2024beta,linot2023dynamics,xia2023hierarchical}. In short, VAEs are neural network models that reduce system dimensionality by learning a nonlinear latent representation of the data via an encoding-decoding architecture. However, traditional VAEs often use Gaussian distributions, limiting their ability to capture extremes.

\par
\begin{figure}[!t]
    \centering
    \includegraphics[width=0.9\linewidth]{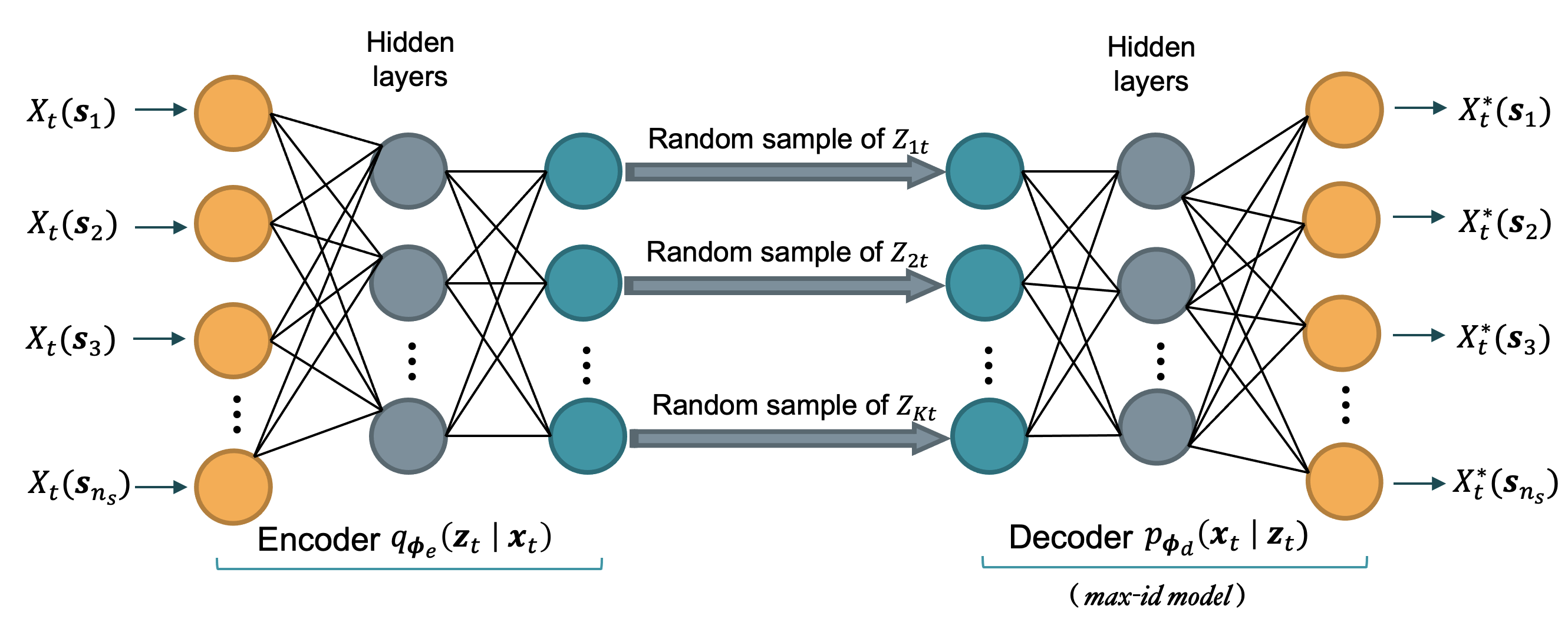}
    \caption{Schematic diagram of the xVAE. This model takes as input the spatial indexed processes $\{X_t(\bs):\bs\in \mathcal{S}\subseteq\mathbb{R}^2\}$ over time $t=1,\ldots, n_t$. A probabilistic encoder compresses these spatiotemporal processes into a set of $K$-dimensional latent variables $\{Z_{1t}, \ldots, Z_{Kt}\}$, which capture essential dynamical and spatial features. Then the decoder neural networks maps these latent representations back to the data space, generating synthetic realizations $\{X^*_t(\bs):\bs\in \mathcal{S}\subseteq\mathbb{R}^2\}$ that preserve the statistical and structural properties of the original input. Here, $\mathcal{S}$ is the spatial domain of interest, and $\bs$ represents any location within that domain. Although this work focuses on $\mathcal{S}\subseteq\mathbb{R}^2$, this framework can be easily extended to three-dimensional domains ($\mathcal{S}\subseteq\mathbb{R}^3$). Further details are provided in Section~\ref{sec:xVAE_description}.} 
    \label{fig:vae_diagram}
\end{figure}

  In this paper, we implement a novel probabilistic statistical framework to explicitly accommodate concurrent and dependent extremes within a VAE.
We embed a new flexible max-infinitely divisible (max-id) process within the VAE’s hierarchical encoding-decoding architecture, creating what we call the xVAE \cite{zhang2023flexible, zhang2024capturing}. In essence, the encoder in the VAE compresses complex spatial dependencies into a simpler latent space of dimension $K$, while the decoder samples from this space to generate new data. Instead of Gaussian variational distributions, xVAEs use heavy-tailed distributions (e.g., exponentially-tilted positive stable distributions), enabling them to give the VAE structure so-called ``max-infinite divisibility" and allow it to better capture extreme events; see Fig.~\ref{fig:vae_diagram} for a schematic illustration and Section \ref{sec:xVAE_description} for details. The tail index parameter offers flexibility in modeling varying levels of extremity, accommodating heterogeneous patterns across space and time. Extensive studies in \cite{zhang2023flexible} highlight xVAE's effectiveness for capturing extreme dependencies in spatial data and its ability to rapidly generate realistic ensemble simulations with uncertainty quantification.

\par The xVAE offers several advantages over VAEs and other methods for dimensional reduction 
as they learn a compact, latent representation that captures the essential structures of  both the bulk and the extremes in the turbulence data. 
The stochastic characteristics of turbulence are well-represented through the max-id model that has the flexible tail dependence properties. While training the xVAE is computationally intensive---like all VAEs---the cost is amortized so that once trained, generating realizations that resemble the statistical properties of the training data, useful for data augmentation. 
As the xVAE retains the spatio-temporal dynamical information of the turbulence data, the realizations significantly reduce the cost for generating reduced dimensional model and for training physics informed  Neural Networks (PINN). The decoder of a trained xVAE can act as a real-time surrogate model for turbulence, predicting flow fields orders of magnitude faster than traditional simulation methods like Direct numerical simulation (DNS) or Large-eddy-simulation (LES).
Overall, xVAE is very powerful and promising to capture extreme turbulent events that combines the strengths of data-driven models and statistical turbulence representations, opening pathways for efficient analysis and better uncertainty quantification (UQ) for risk assessment across diverse scientific and engineering domains.

The methodology is demonstrated on a wildland turbulent fire plume at very high Reynolds number, where extreme, intermittent bursts of Turbulent Kinetic Energy (TKE) arise from buoyancy-driven dynamics at the fire source. Turbulent plumes pose a significant challenge due to the complex nonlinear coupling between thermodynamic processes and velocity fields, which drives turbulent production and results in extreme fluctuations \cite{chen2023energetics,bhaganagar2020numerical,chen2021new}. These plumes are highly relevant in meteorological contexts, such as convective plumes driven by diurnal heating or wildland fire events. Plume dynamics also play a crucial role in the atmospheric transport of hazardous or contaminant gases, whether originating from natural extreme events \cite{britter1989atmospheric} or anthropogenic releases \cite{bhaganagar2017assessment, bhaganagar2020local}. For this work, high-resolution spatial inputs (i.e., scaled fluid density) have been generated using large-eddy simulations (LES) that have been rigorously validated against physical principles and prior observations \cite{chen2023energetics}. We will apply both the state-of-the-art POD and xVAE to emulate/reconstruct these LES inputs at the $x$-$z$ plane of a fixed $y$ level.

\subsection{Large-Eddy-Simulation of a Turbulent Plume}\label{sec:LES}

The turbulent plume was generated using high-resolution LES methodology---bplume-WRF-LES developed by Bhaganagar and BhimiReddy \cite{bhaganagar2021MWR}---which has been modified within the Weather Research Forecast framework (WRF-ARW v4.1) \cite[hereafter WRF;][]{skamarock2008description}.

Within this framework, the WRF-LES model is coupled with an active scalar transport equations, which were specifically developed and validated for simulating buoyant plumes \cite{bhaganagar2020numerical, bhaganagar2021MWR}.  Specifically, the WRF-LES model generates the turbulent buoyant plume from an axisymmetric heated source of diameter $D$ into a quiescent environment; see \cite{bhaganagar2020numerical} for details of the WRF-LES methodology. For the sake of completeness, the mathematical framework is briefly described in Section~\ref{appen:wrf-les} in the Supplementary Material.

This WRF-LES model was used to simulate the September 2022 Mosquito wildland fire (California's largest wildland fire in 2022). Realistic boundary conditions for the simulation were obtained from NOAA’s High-Resolution Rapid Refresh (HRRR) model. A heated air plume was continuously released from a circular ground-level source with a diameter of 400m, centered at  ($39.006^\circ$N, $120.745^\circ$W). The background atmospheric velocity profile and temperature profile, as well as surface heat flux, were obtained from the WRF-LES simulations initiated at 11am on 09/09, in which buoyancy originates from \textit{density differences} between the lighter source gas and the ambient air. In this work, the variable of interest is the non-dimensionalized density of the fluid scaled by the density at the source:
\begin{equation}\label{eqn:def}
    X_t(\bs)=\frac{\rho_t(\bs)-\rho_a}{\rho_s-\rho_a},
\end{equation}
in which $\rho_a$ is the ambient fluid density and $\rho_s$ is the source fluid density. When buoyant hot gases rise from a fire plume, the gas density $\rho_t(\bs)$ near the top of the plume could be lower than the ambient air density, leading to negative values of $X_t(\bs)$ at time $t$.

The LES domain computational domain spanned 4km $\times$ 4km $\times$ 7km (i.e., 10D $\times$ 10D $\times$ 17.5D) along the two cross-stream (horizontal) and axial (vertical) directions, with the source located at the center of the bottom boundary. A uniform Cartesian grid discretized the domain into 100 $\times$ 100 $\times$ 700 nodes in the cross-stream and axial directions. Periodic boundary conditions were imposed on the side boundaries, whereas a constant-pressure boundary condition was imposed at the top. At the source, a constant buoyancy flux of $1.07 \times 10^4 m^4 s^{-3}$ was prescribed, with no momentum contribution, so the flow was driven purely by buoyancy forces. Simulations ran for 50 minutes with data recorded every 30 seconds (i.e., $n_t=100$ times in total). Planar data were extracted from the grid locations on the $x$-$z$ plane fixed at $y=50$ (i.e., $2$ km), the centerline of the plume. The number of grid locations on this $x$-$z$ plane is $n_s=70,000$.

\section{Results}\label{sec:results}

\subsection{Optimize Latent Dimensionality for Accurate Emulation}\label{sec:cv}
Both POD and xVAE reduce the dimensionality of the original spatial input at a given time (denoted by $n_s$) to a compact latent space of dimension $K$ (see equations~\eqref{eqn:POD_reduce} and \eqref{eqn:low_rank_representation} for POD and xVAE respectively).
Selecting an appropriate number of basis functions $K$ is crucial for both the POD and the xVAE, as it balances model complexity and computational efficiency while preserving essential spatial structures. 

To determine the optimal $K$, we perform cross-validation by dividing the $100$ time replicates into $10$ folds, holding out one fold at a time for validation while training the POD or xVAE model on the remaining 9 folds. We then evaluate emulation performance across the holdout folds using a key metric. 

The metric we are using is based on the Root Mean Squared Error (RMSE), which is a valuable metric for evaluating the fidelity between the data inputs and their emulation and considers the entire distribution. However, RMSE achieves zero when the number of basis functions equals the number of time replicates $n_t$ in POD, indicating $K=n_t$ is always the best choice based RMSE because there is perfect preservation of variation in this case. Similarly for xVAE, $K=n_t$ will also lead to the best RMSE but incurs higher computational cost. To mitigate this, we focus on the tail similarity between the observations and the emulations and examine the tail RMSE, defined as:
\begin{equation}
    \text{tailRMSE}(p)=\sqrt{\frac{1}{n_tn_s}\sum_{t=1}^{n_t}\sum_{i=1}^{n_s} \{X_t(\bs_i)-X^*_t(\bs_i)\}^2 \mathbbm{1}\{X_t(\bs_i)< x^{p}\}},
\end{equation}
in which $\mathbbm{1}\{x< x^{p}\}$ is an indicator function of $x\in\mathbb{R}$ defined as 
\begin{equation*}
    \mathbbm{1}\{x< x^{p}\}=\begin{cases}
        1,&\text{ if } x< x^{p},\\
        0,&\text{ if } x\geq x^{p},
    \end{cases}
\end{equation*}
and $p$ is a small quantile level close to zero with $x^{p}$ representing a low threshold set at the $p$th quantile of the observations across all times and locations. This focus on low extremes is particularly relevant, as low extremes in $\{X_t(\bs)\}$ correspond to the strong buoyancy or intense turbulent mixing where the fluid is significantly lighter than the surrounding environment; see equation~\eqref{eqn:def}. If the interest is instead in high extremes, the indicator function can be modified accordingly to censor the highest values rather than the lowest.

\begin{figure}[!t]
    \centering
    \includegraphics[width=0.45\linewidth]{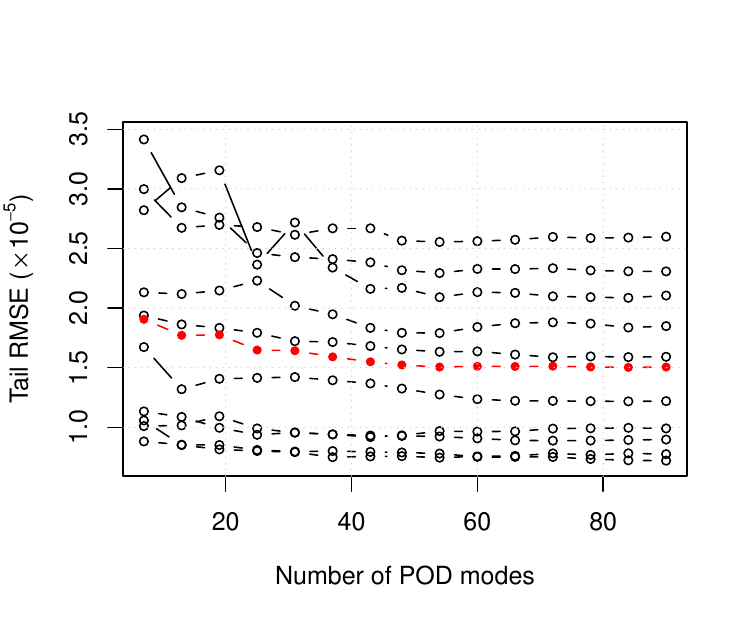}
    \includegraphics[width=0.45\linewidth]{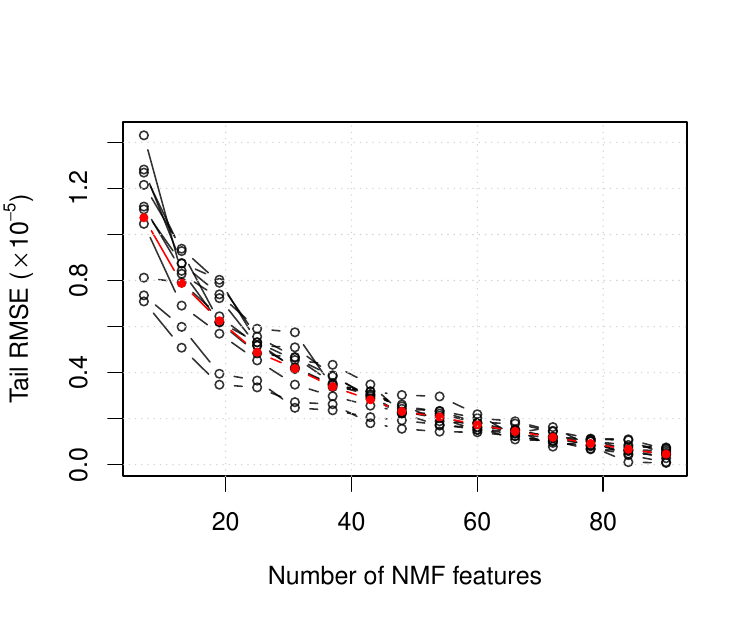}
    \caption{The tail RMSE values with $p=0.05$ from the 10-fold cross-validation are plotted for the number of bases $K$ ranging from 7 to 90. The red dotted line represents the average tail RMSE across all 10 folds. For clarity, the average tail RMSE obtained from the xVAE model using 50 bases across all folds is $2.19\times 10^{-6}$, highlighting the much lower converged level of the tail RMSE.}
    \label{fig:POD_CV}
\end{figure}

In POD, prediction involves projecting holdout time replicates onto the subspace spanned by principal components (see Fig.~\ref{fig:POD_CV}, left panel). We examine a varying number of principal components used to perform the prediction ranging from 7 to 90. Since there are only 90 time replicates in each training of POD, the last number corresponds to the maximum information that exists in the training set. Tail RMSE at $p=0.05$ generally decreases with increasing number of modes, stabilizing around 0.001 after 50 components when averaged across folds (red line in Fig.~\ref{fig:POD_CV}). Variability in prediction performance exists across folds, with the first fold showing the highest tail RMSE, largely due to differing scales between training and holdout sets. Since the scaled density of the plume increase from $-1\times 10^{-3}$ at time 1 to $-1\times 10^{-4}$ at time 100, POD has difficulty in predicting the scales correctly when the holdout set of fold 1 has larger scales than the training set.

For xVAE, weights and biases of neural networks are learned from the training set, and then holdout times are inputted into the encoder-decoder to obtain emulations. The major advantage of xVAE over POD is that we can rapidly emulate 1,000 replicates of each hold time, and then we average the tail RMSEs calculated from each replicate. xVAE consistently outperforms POD (Fig.~\ref{fig:POD_CV}, right panel), with less variability in tail RMSE values across folds converging to $10^{-5}$. This indicates xVAE's robustness in retaining extremal dependence and being less sensitive to scale differences in the holdout set. However, initial folds still show less accuracy compared to later folds.

\begin{figure}[!t]
    \centering
    \includegraphics[height=0.35\linewidth]{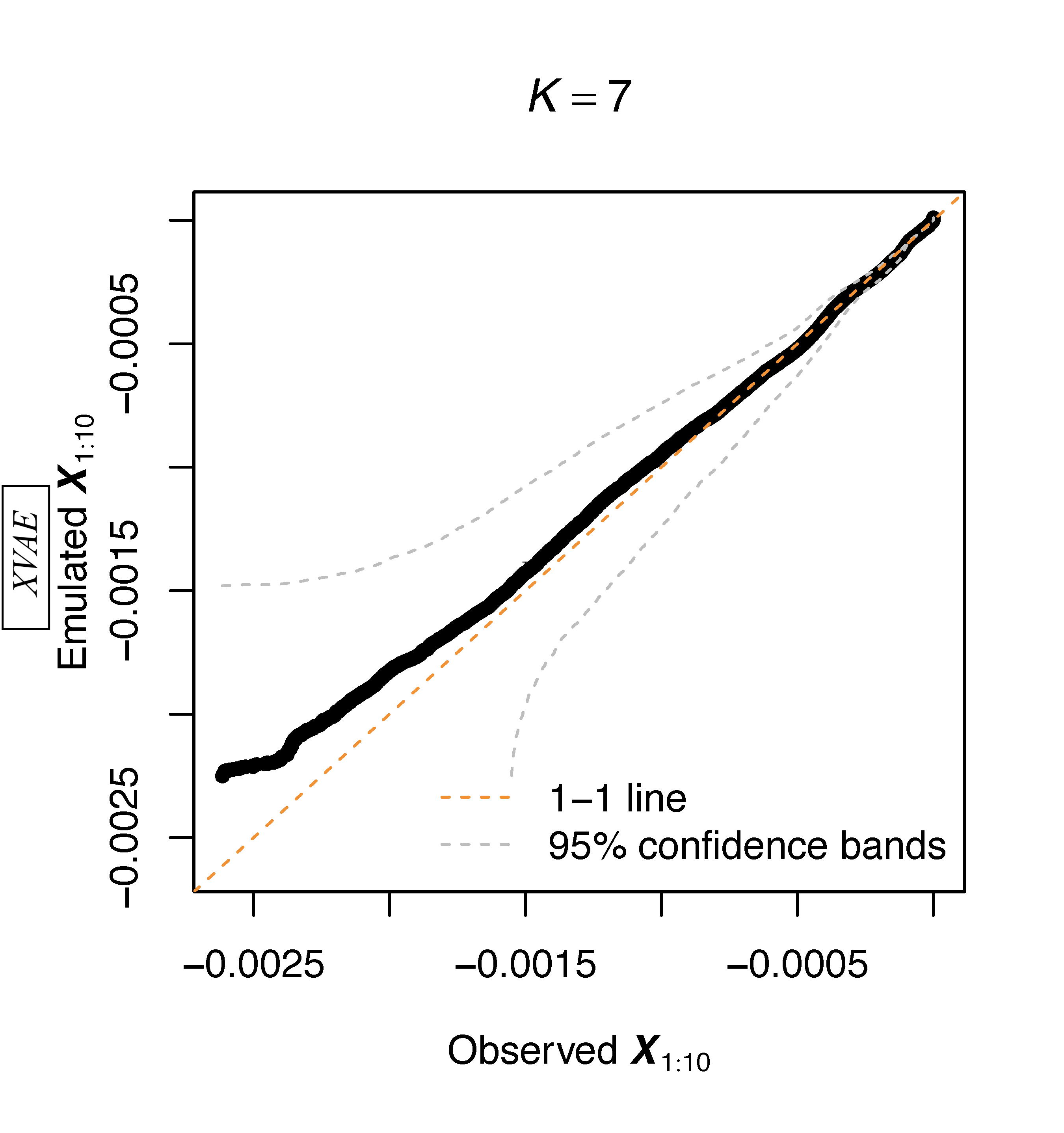}
    \includegraphics[height=0.35\linewidth, trim={1cm 0 0 0}, clip]{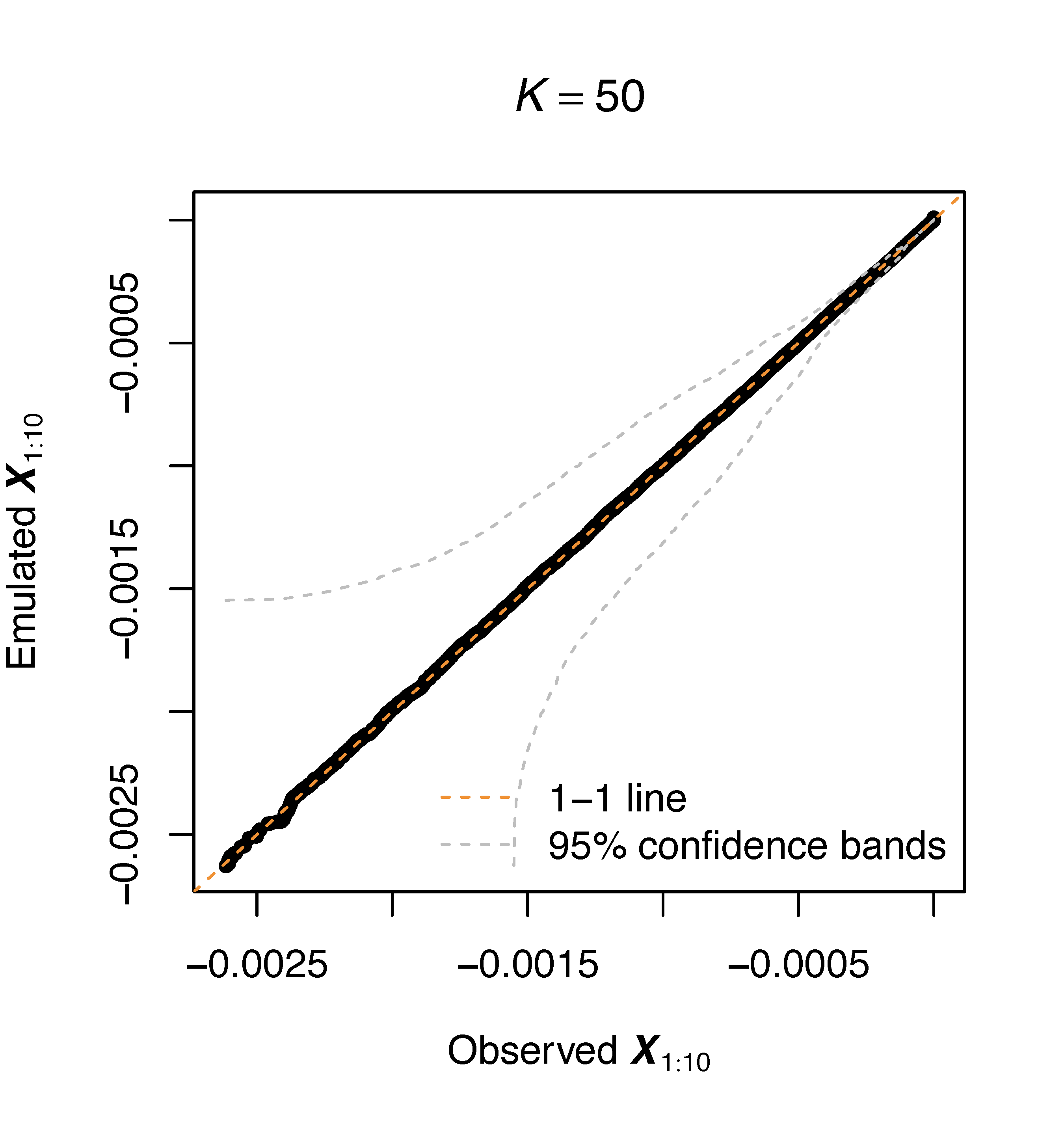}
    \includegraphics[height=0.35\linewidth, trim={1cm 0 0 0}, clip]{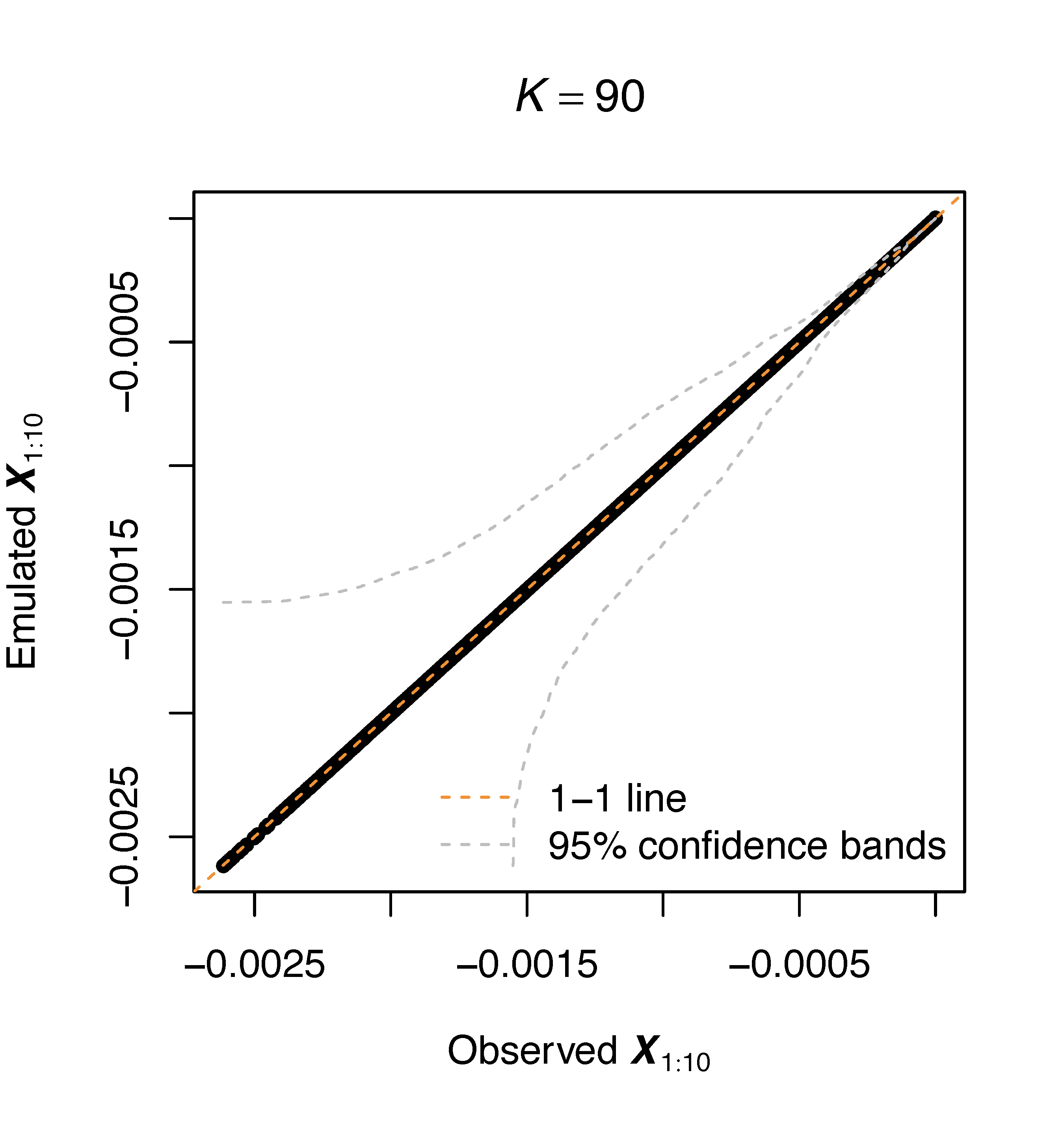}

    \includegraphics[height=0.35\linewidth]{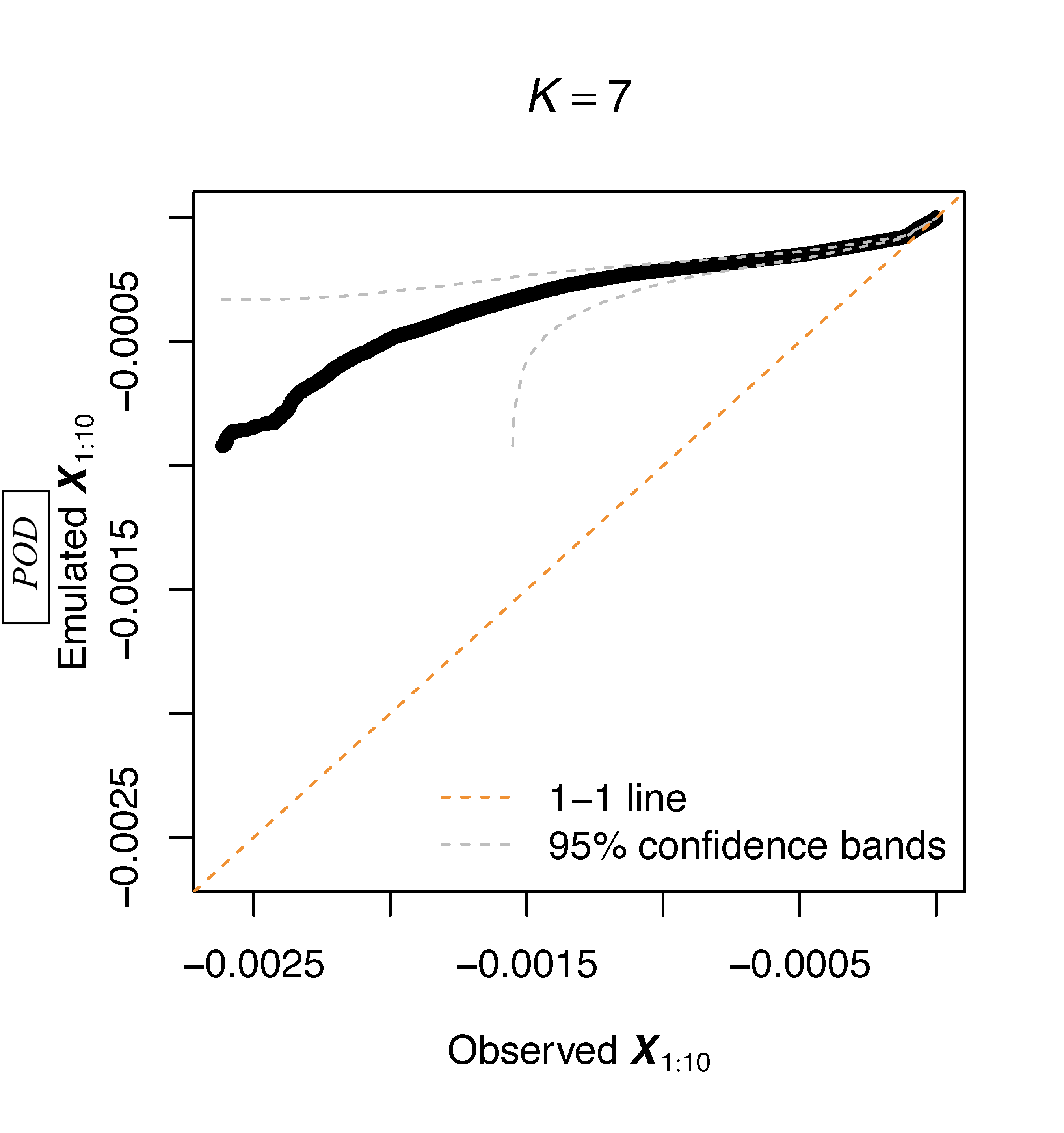}
    \includegraphics[height=0.35\linewidth, trim={1cm 0 0 0}, clip]{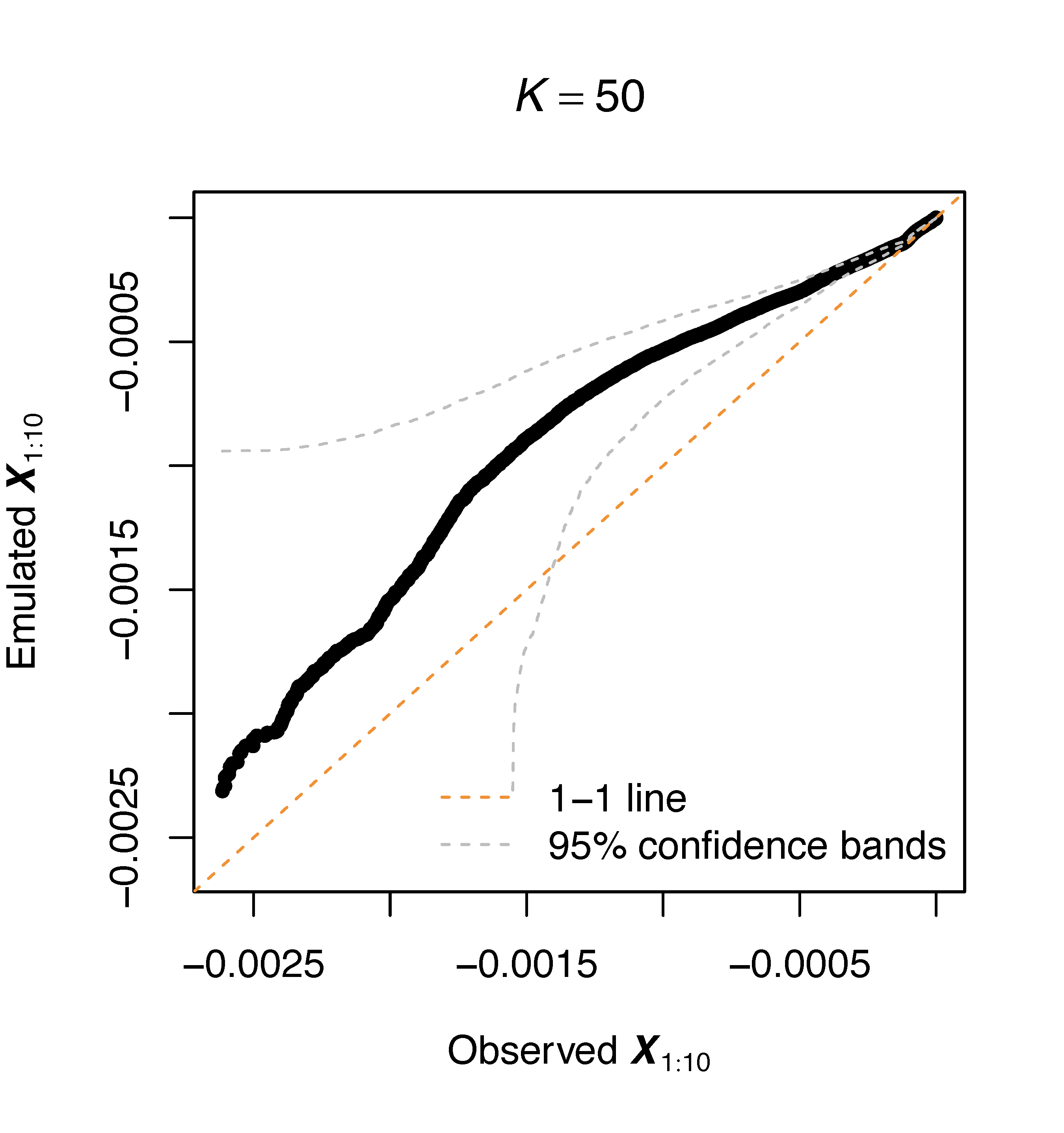}
    \includegraphics[height=0.35\linewidth, trim={1cm 0 0 0}, clip]{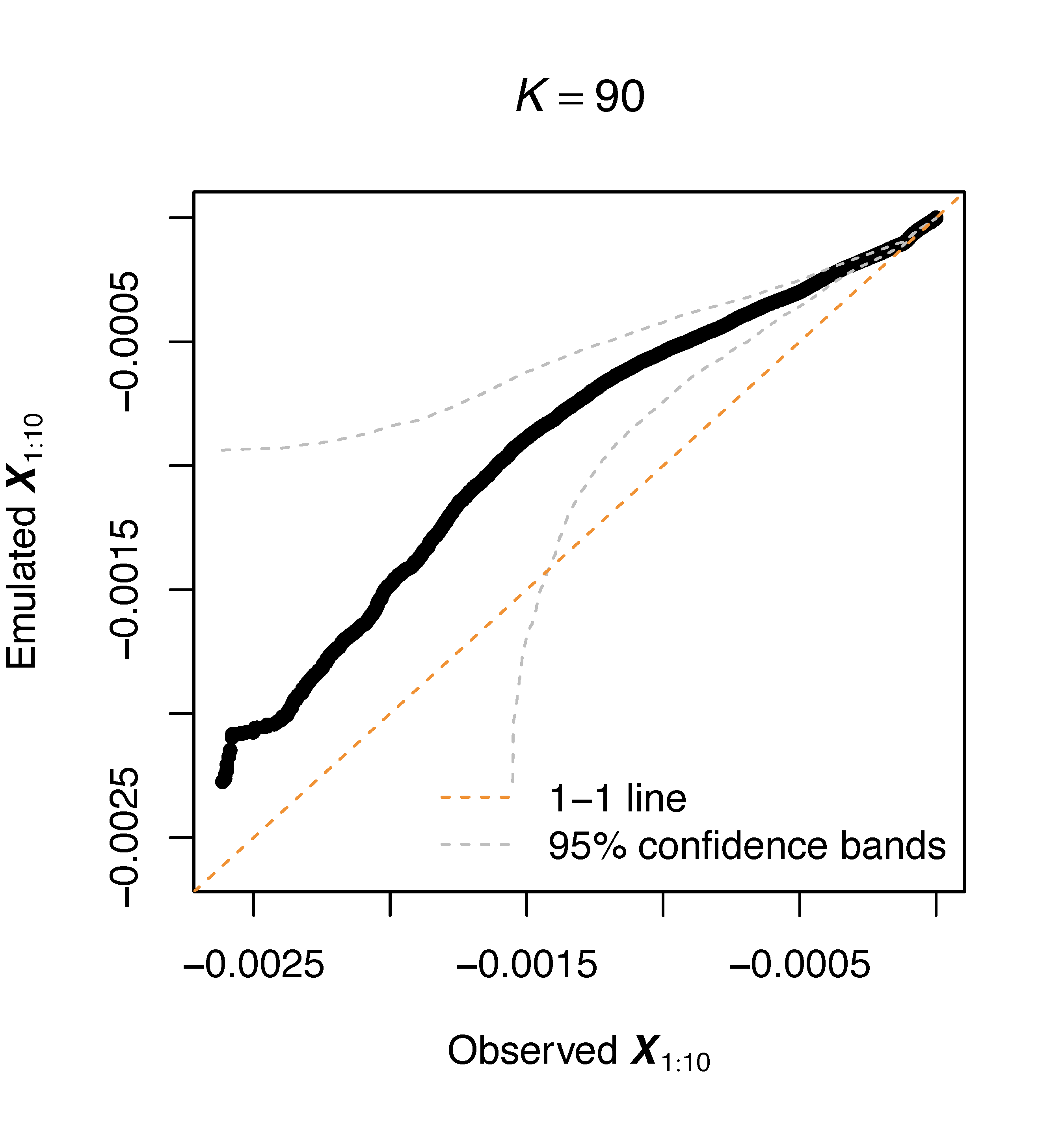}
    \caption{QQ-plots comparing the held-out observed data from fold 1 (i.e., times 1 to 10 are excluded) with the corresponding emulated fields using xVAE (top panels) and POD (bottom panels). The plots illustrate various numbers of bases: $K=7$ (left), $K=50$ (middle), and $K=90$ (right). For similar plots corresponding to fold 10, where times 91 to 100 are held out, see Fig.~\ref{fig:qqplots2} in the Supplementary Material.}
    \label{fig:qqplots1}
\end{figure}

Additionally, Fig.~\ref{fig:qqplots1} presents the quantile-quantile (QQ) plots by pooling observed and emulated spatial data across times 1 to 10 into the same plot. This comparison assesses whether the spatial input and emulation exhibit similar ranges and quantiles for the first fold, where the initial 10 times were excluded from the training of xVAE and POD. Although data pooling across space and time ignores spatial dependence and temporal non-stationarity, QQ-plots still provide value in determining whether the spatial distribution is similar at all quantile levels. The top panels indicate that xVAE’s emulation performance stabilizes once $K$ reaches 50, while the bottom panels reveal that POD struggles to accurately capture the scale of extreme values. Notably, the overall quantile level at $p=0.001$ of all scaled densities across times 1 to 10 is approximately $-0.0018$. The deviations from the 1-1 line seen in the left panels of Fig.~\ref{fig:qqplots1} arise from the lowest 0.1\% of all observations. Generally, POD effectively captures the bulk of the spatial distribution (i.e., observations much greater than $-0.0018$). However, even with $K=90$, POD tends to misrepresent the tail dependence strength, as evidenced by high tail RMSE values and an overestimation of extreme values, as shown in the QQ plots. While holding out times is not typically practiced in POD analyses of plume turbulence data, the results in this subsection provide an objective method for selecting an optimal number of basis functions for both xVAE and POD. Moreover, they highlight that xVAE is more adept at capturing tail dependence when using trained basis functions from existing times to emulate new observations in an online learning context. Fig.~\ref{fig:qqplots2} in the Supplementary Material shows similar QQ-plots examining the emulation performance for the 10th fold.

In subsequent comparisons, we present emulation results from training on the entire $K=100$ time replicates using 50 bases using both POD and xVAE, as we see an elbow change when the number of bases increases to 50 for both methods.


\subsection{Model evaluations}\label{sec:res}
\begin{figure}
    \centering
    \includegraphics[width=1.05\linewidth]{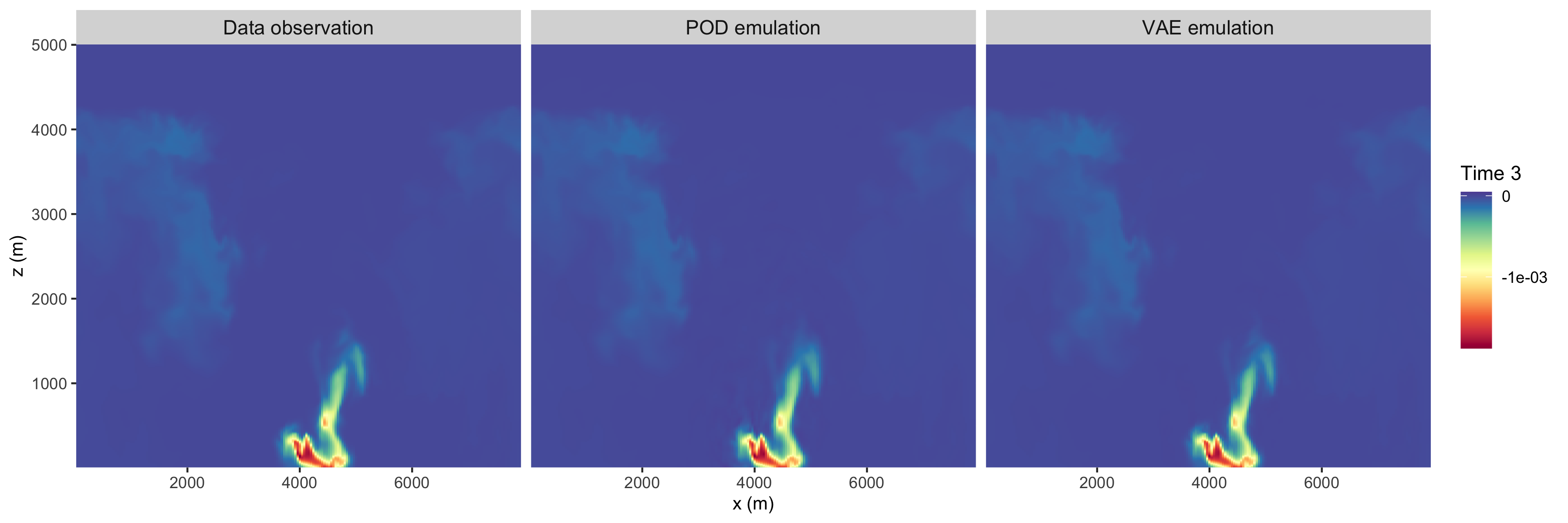}
    \includegraphics[width=1.05\linewidth]{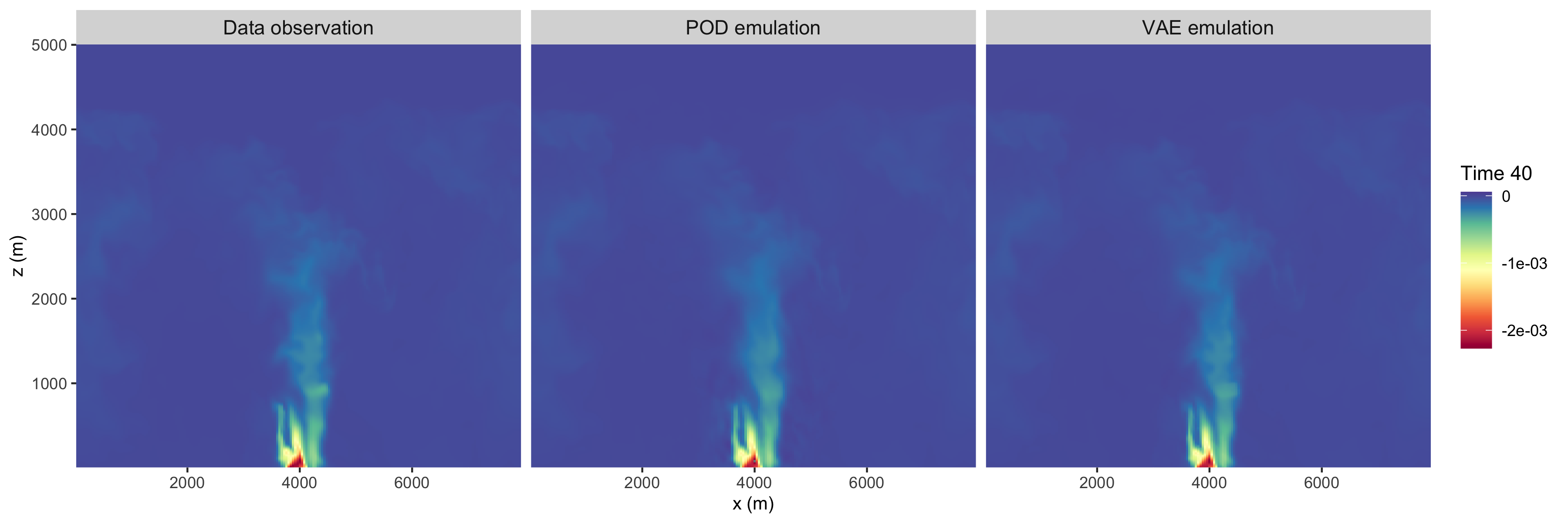}
    \includegraphics[width=1.05\linewidth]{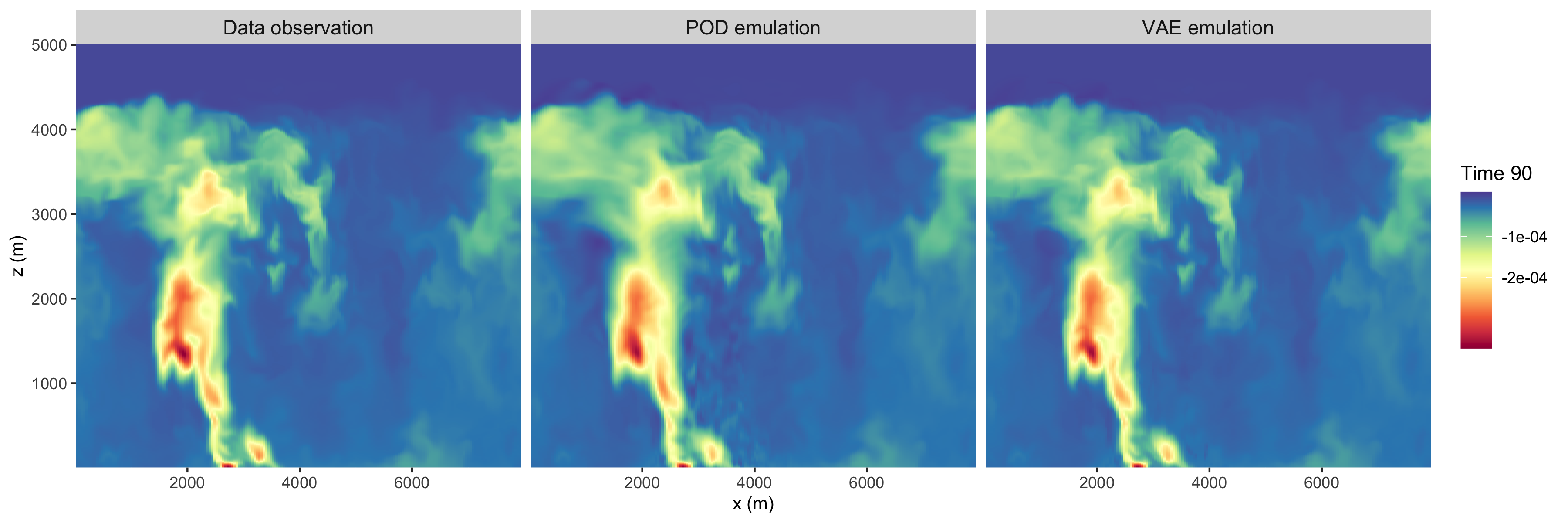}
    \caption{Scaled fluid density fields generated using the bplume-WRF-LES model (left) and its corresponding emulated fields at three randomly selected times. Note the data scales are different in different rows to show more spatial structures.}
    \label{fig:emulation-res}
\end{figure}
Fig.~\ref{fig:emulation-res} displays emulated replicates from xVAE and POD, both trained with all 100 time replicates. For comparison, we examine the emulations against turbulence plume density flows at three distinct times. Generally, xVAE and POD closely replicate the spatial patterns of turbulence across these times. Although the POD emulation at time 90 exhibits some oversmoothing, visually distinguishing which emulator performs better is challenging. The tail RMSE calculations reveal that xVAE has a tail RMSE of $5.32\times 10^{-6}$ at $p=0.05$, while POD has a tail RMSE of $9.68\times 10^{-6}$. Compared to the results from cross-validation shown in Fig.~\ref{fig:POD_CV}, POD shows greater improvement in tail RMSE when trained without data exclusion. 

Neither tail RMSE nor QQ plot directly evaluates the extremal dependence, i.e., the simultaneous occurrence of extreme values. In the spatial extremes literature, extremal dependence at two locations $\bs_i$ and $\bs_j$ is commonly described by the measure 
\begin{equation}\label{eqn:chi}
  \chi_{ij}(p) = \Pr\{F_j(X_j) > p \mid F_i(X_i) > p\} = \frac{\Pr\{F_j(X_j) > p, F_i(X_i) > p\}}{\Pr\{F_i(X_i) > p\}}\in [0,1],
\end{equation}
in which $p\in(0,1)$ and $F_i$ and $F_j$ are the continuous marginal distribution functions for $X_i\equiv X(\bs_i)$ and $X_j\equiv X(\bs_j)$, respectively; see \cite{davison2015statistics} for example. When $p$ is close to one, $\chi_{ij}(p)$ measure the probability that one variable is simultaneously extreme given that the other variable is similarly extreme.

Following \cite{zhang2023flexible}, we estimate $\chi_{ij}(p)$ empirically by treating $\{X_t(\bs)\}$ as a stationary and isotropic spatial process at each time $t$, and assuming independent distribution across time. This independence assumption across time provides multiple replications for the empirical estimates of $\chi_{ij}(p)$. By assuming stationarity and isotropy, we simplify $\chi_{ij}(p)$ to $\chi_{d}(p)$, with $d=||\bs_i-\bs_j||$ denoting the distance between locations. Although the plume turbulence data is neither stationary nor isotropic, making $\chi_{d}(p)$ a partial characterization of the dependence structure, it still serves as a useful summary statistic and provides insight into the average dependence decay with distance, regardless of direction.

The metric $\chi_{ij}(p)$ is empirically estimated for each data set (observed or emulated) as follows. Firstly, the marginal distribution function $F_i$ at any location $\bs_i$ is estimated by the empirical distribution function $\hat{F}_i$ of all records across time at $\bs_i$. For a fixed $d$, we find all pairs of locations with similar distances (within a small tolerance, say $\epsilon=0.001$), and compute the empirical conditional probabilities $\widehat{\chi}_d(p)$ at a grid of $p$ values. Confidence envelopes are determined by treating the outcome (i.e., whether or not both locations simultaneously exceed $p$) for each pair as a Bernoulli variable and computing pointwise binomial confidence intervals, assuming the independence of all pairs. 

\begin{figure}[!t]
    \centering
    \includegraphics[height=0.43\linewidth]{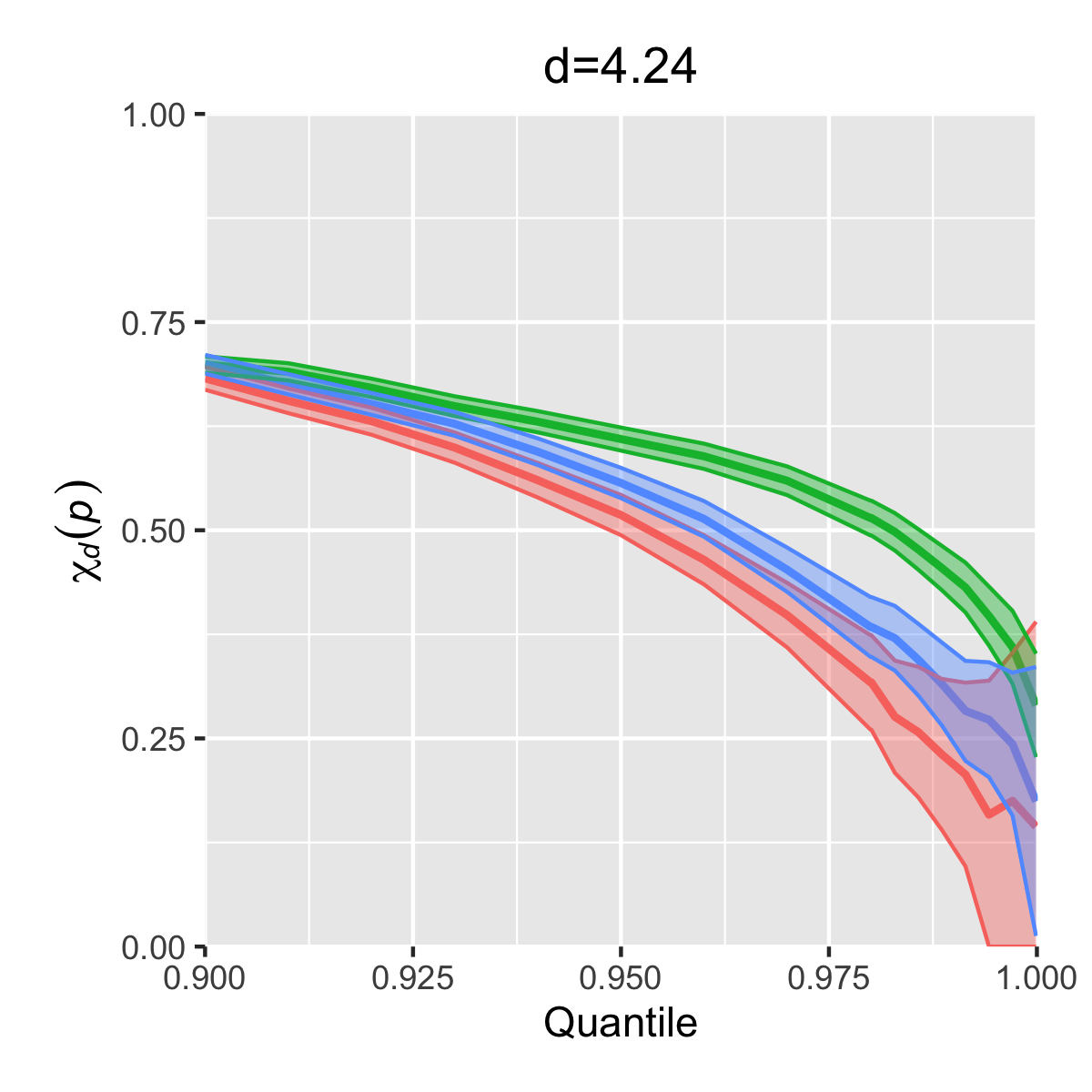}
    \includegraphics[height=0.43\linewidth]{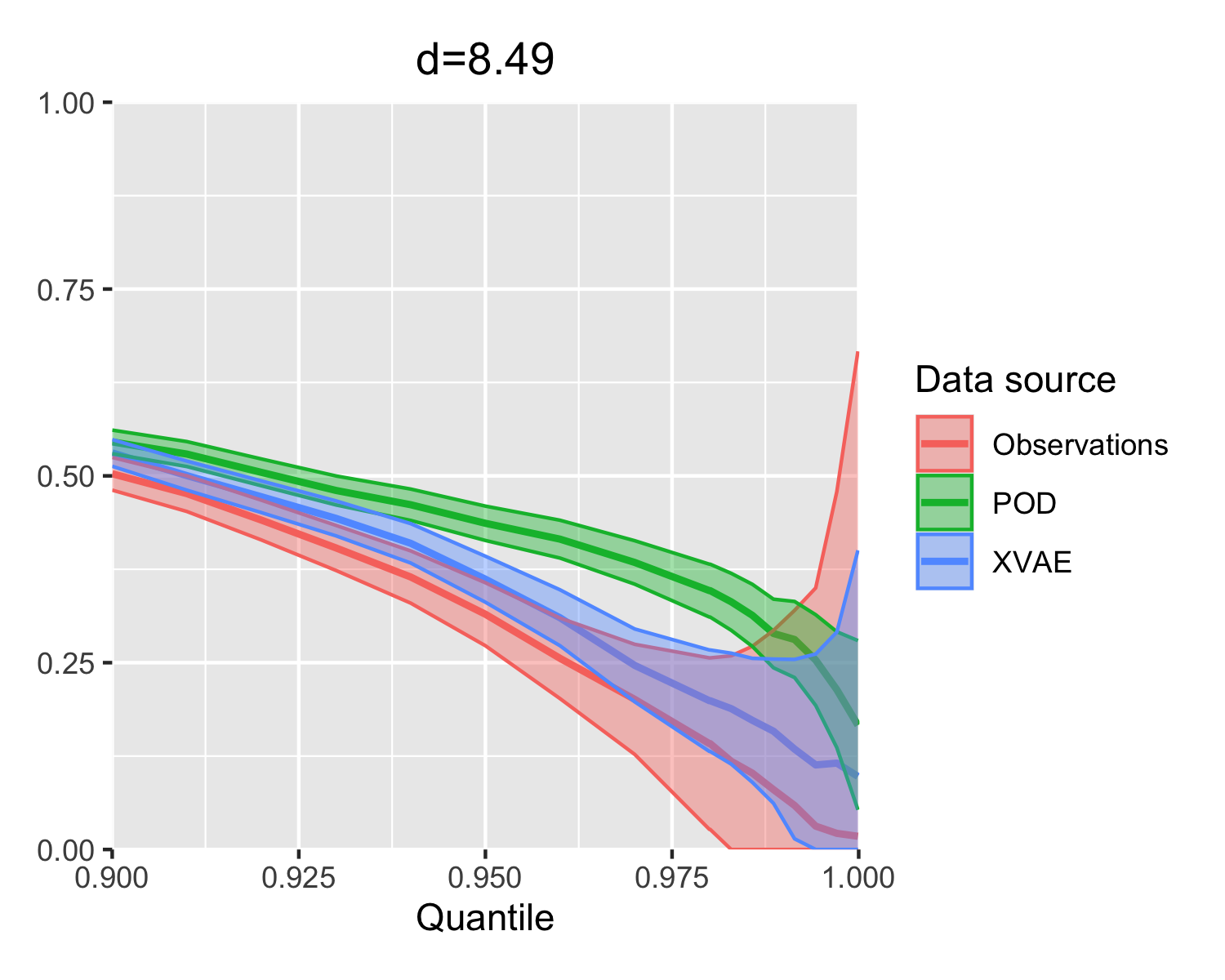}
    \caption{We show the empirically-estimated $\chi_d(p)$ at $d = 4.24$ (left) and $d=8.49$ (right) based on plume observations (red), POD reconstructed data (green) and xVAE emulated data (blue). }
    \label{fig:chi_plots}
\end{figure}
Fig.~\ref{fig:chi_plots} compares the empirical measure $\widehat{\chi}_d(p)$ for $p\in (0.9, 0.9999)$ at $d=3\sqrt{2}\approx 4.24$ and $d=6\sqrt{2}\approx 8.49$ for the observed turbulent buoyant plume and the emulations from POD and xVAE across all times. 
The uncertainty in the empirical estimates increases with the extremity of $p$, and the dependence strength decreases as the pairwise $d$ increases. Fig.~\ref{fig:chi_plots} demonstrates that the dependence measure aligns well between the observed data and the xVAE emulated data at both short and long distances. However, POD tends to overestimate extremal dependence strength at sub-asymptotic levels, i.e., $p\in (0.93, 0.99)$, likely due to oversmoothing. This is consistent with the QQ-plots in Fig.~\ref{fig:qqplots1}, which reveal mischaracterizations of marginal scales. The overestimation of dependence in POD further indicates difficulties in capturing joint tail behavior, suggesting that xVAE should be preferred for accurately assessing spatial extremes in turbulence plumes.

Additionally, Fig.~\ref{fig:chi_plots} illustrates a weakening of dependence as the quantile level $p$ increases for a fixed distance $d$. A key aspect of spatial extremes modeling is developing flexible copulas that can exhibit this weakening dependence structure \cite[see][for a detailed review]{huser2022advances}. For instance, Gaussian processes often exhibit a $\chi_{d}(p)$ that converges to $0$ as $p\rightarrow 1$ at a very fast rate unless $d=0$ or the underlying correlation equal to 1, which max-stable processes have a rigid $\chi_{d}(p)$ that becomes constant for high $p$ levels. Although Gaussian processes are most commonly used for spatial emulation, \cite{zhang2023flexible} demonstrated that they struggle to emulate processes with extreme values in their extensive simulations. A copula with a flexible dependence structure is thus crucial for accurately reflecting the dependence structure of the observed data. Fig.~\ref{fig:chi_plots} highlights the effectiveness of the flexible max-id copula employed in xVAE, whereas POD mischaracterizes the dependence structure, even when using data-derived bases. 

\begin{figure}[!t]
    \centering
    \includegraphics[height=0.43\linewidth]{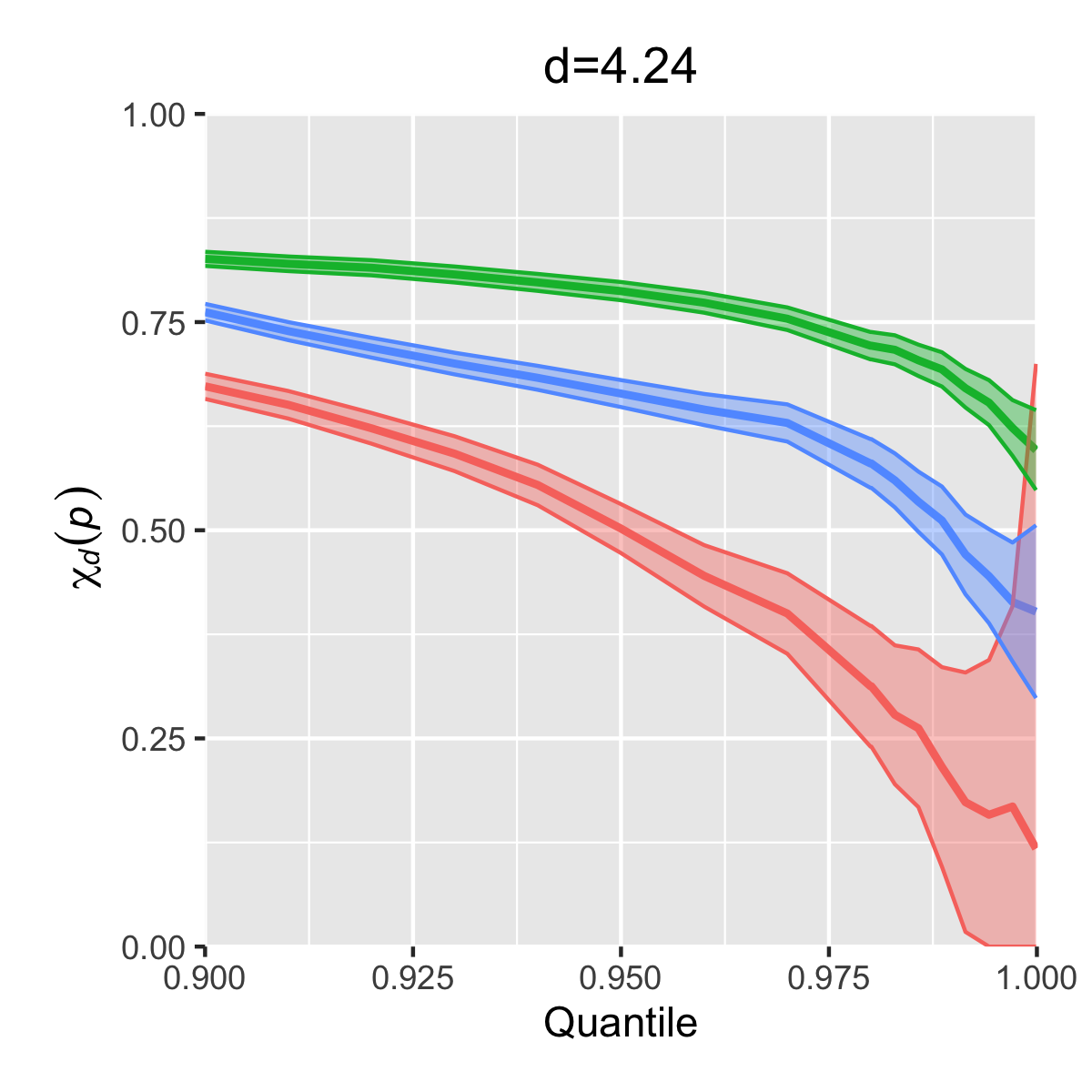}
    \includegraphics[height=0.43\linewidth]{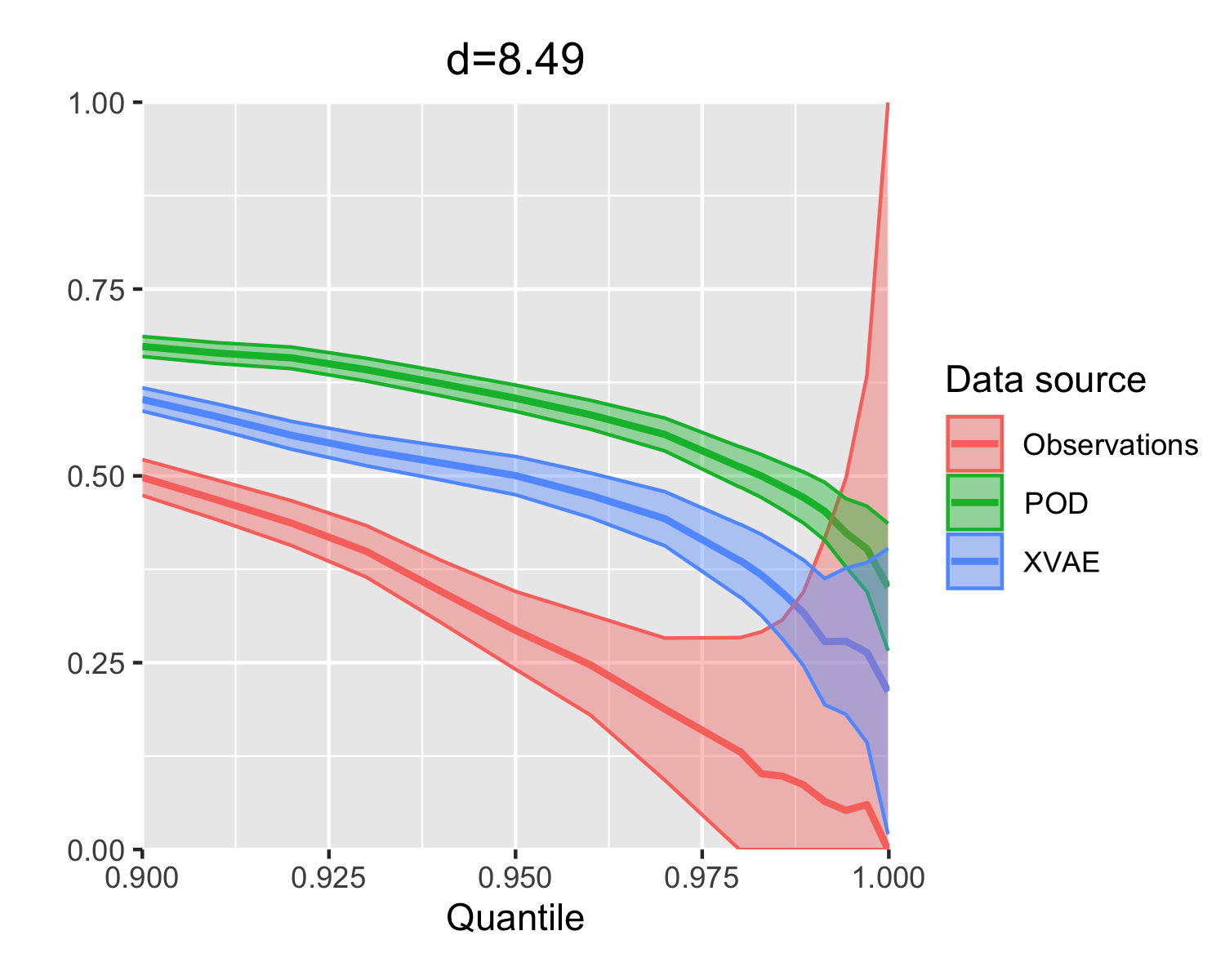}
    \caption{Using 7 bases, we show the empirically-estimated $\chi_d(p)$ at $d = 4.24$ (left) and $d=8.49$ (right) based on plume observations (red), POD reconstructed data (green) and xVAE emulated data (blue). }
    \label{fig:chi_plots_ncomp7}
\end{figure}
As a side note, we aim to highlight the risks associated with using too few bases in the emulation of plume turbulence. As previously mentioned, it is common practice to use 3--7 POD bases for rapid computations in plume turbulence analysis. Figs.~\ref{fig:POD_CV} and \ref{fig:qqplots1} already show that the marginal scales are not captured correctly when the number of bases $K$ is not sufficient. Fig.~\ref{fig:chi_plots_ncomp7} further illustrates this issue by showing the empirical dependence measure for emulations from POD and xVAE using only $K=7$ bases (principal component bases for POD and non-negative matrix factorization bases for xVAE). Both POD and xVAE overestimate dependence strength in this scenario, though xVAE performs slightly better. Figs.~\ref{fig:chi_plots_cv} and \ref{fig:chi_plots_cv_holdout} in the Supplementary Material show similar results to Figs.~\ref{fig:chi_plots} and \ref{fig:chi_plots_ncomp7}, but with the first ten times held out for the training of POD and xVAE.

\subsection{Uncertainty quantification}\label{sec:UQ}
One major advantage of xVAE over many existing emulators is its ability to effectively quantify uncertainty for various inference targets and assess risks associated with rare events, such as joint high-threshold exceedances. For examples of how UQ can be beneficial, see Section 5 of \cite{zhang2023flexible}. xVAE exemplifies ``amortized inference," where there is a significant upfront training cost. However, once trained, the xVAE can rapidly generate posterior simulations of dependence parameters for the max-id model and synthetic data through the encoding-decoding VAE framework. Efficiently simulating more turbulence data is also useful for downstream analyses

\begin{figure}[!t]
    \centering
    \includegraphics[height=0.25\linewidth]{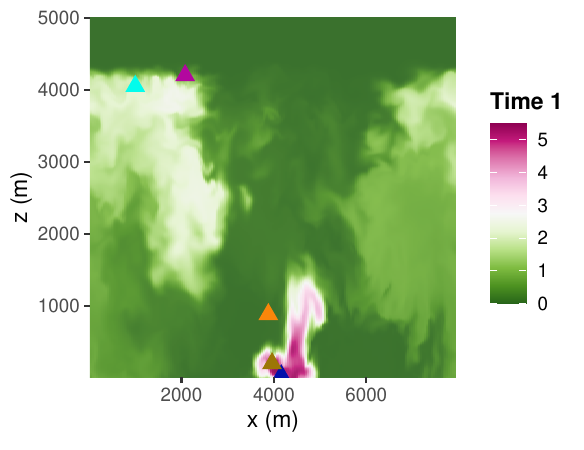}
    \includegraphics[height=0.25\linewidth]{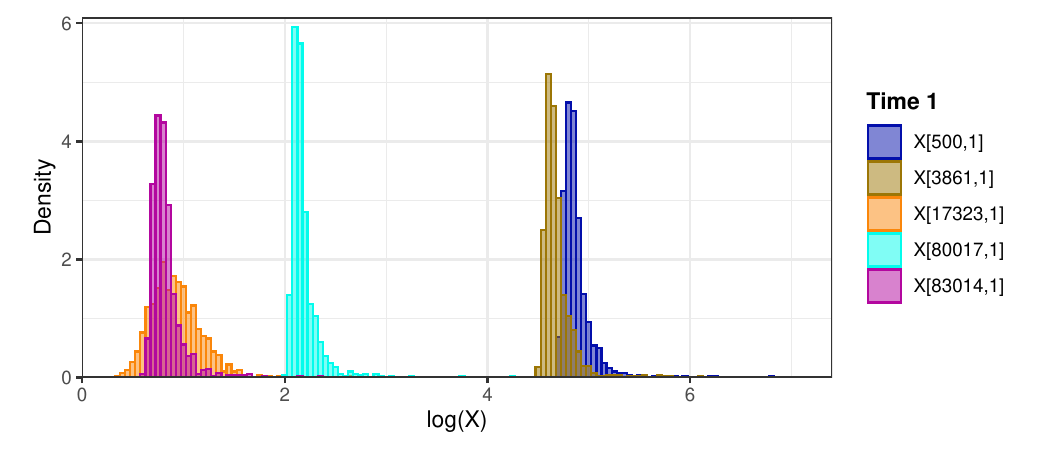}

    \includegraphics[height=0.25\linewidth]{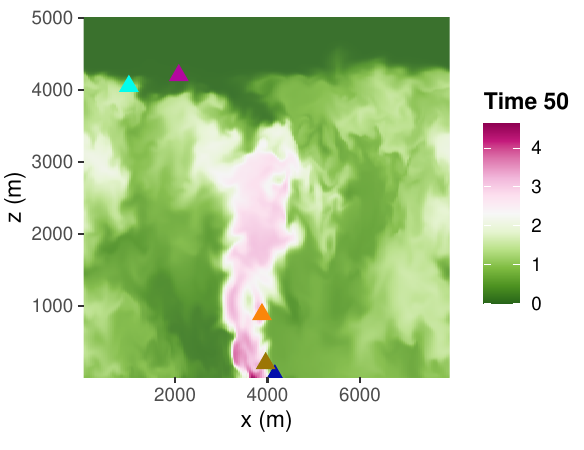}
    \includegraphics[height=0.25\linewidth]{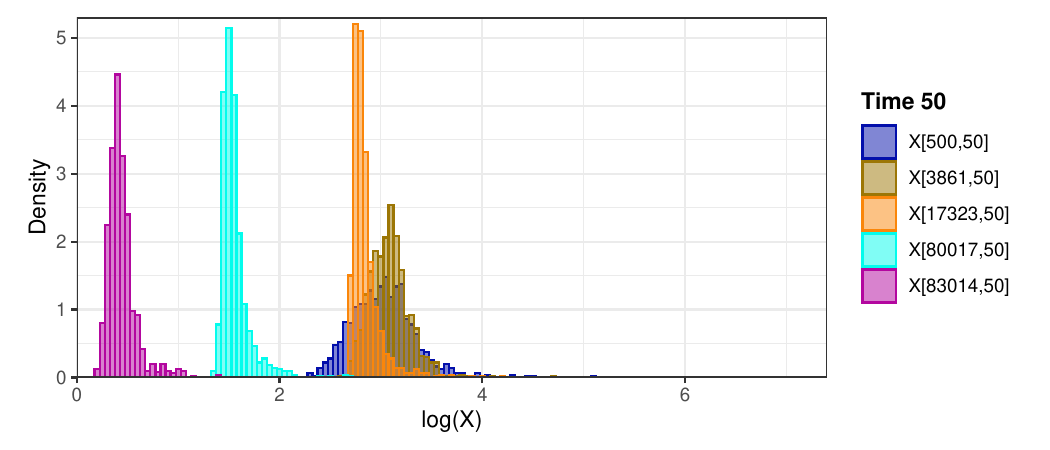}

    \includegraphics[height=0.25\linewidth]{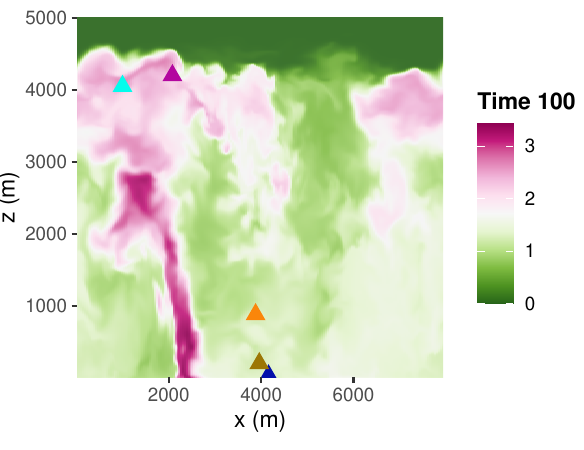}
    \includegraphics[height=0.25\linewidth]{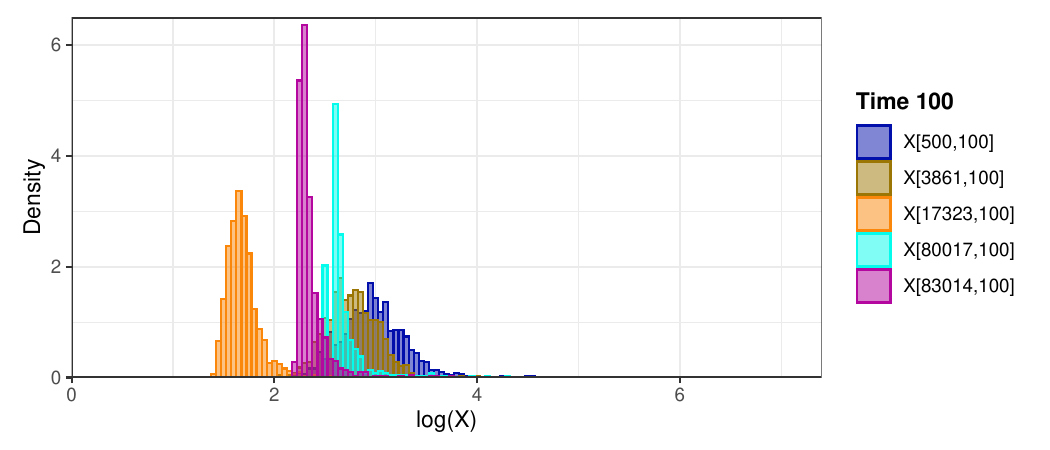}
    \caption{The left panels show the log scale of negated $\bX_1$, $\bX_{50}$ and $\bX_{100}$, respectively. 
    The right panels show the histograms based on 1,000 samples from the emulation. We present density histograms individually for 5 specific sites at 3 times. The locations of these sites are marked in the left panels using the same color scheme as the legend for the histograms.}
    \label{fig:ind_distr}
\end{figure}
For notational simplicity, let $X_{it}=X_t(\bs_i)$ for $i=1,\ldots, n_s$ and $\bX_t = (X_{1t}, \ldots, X_{n_st})^{\rm T}$ for $t=1,\ldots, n_t$. To investigate how the marginal distributions at different locations evolve over time, we generate 1,000 emulations of $\{\bX_t:t=1, \ldots, n_t\}$ using xVAE. For each location and time point, we thus obtain 1,000 replicates to assess variational Bayes-based marginal uncertainty. Fig.~\ref{fig:ind_distr} shows the marginal distributions of $\log(X_{it})$ for $i = 500$, $3861$, $17323$, $80017$, $83014$ at $t=1,\; 50,\;100$. The plots reveal how the density shapes change over time at each location. For instance, the left panels illustrate that the plume does not reach location $\bs_{17323}$ at time 1 but passes through it at time 50, and extends beyond it at time 100. This shift is evident in the orange histograms of the right panels, which show a shift in scale. Additionally, the plume intensity at $\bs_{17323}$ becomes more heavy-tailed at time 50, whereas it is more spread out at times 1 and 100. Such variations in the shape and scale of marginal distributions are not captured by existing emulators, including POD. Furthermore, densities at locations $\bs_{500}$ and $\bs_{3861}$ which are very close to each other, overlap significantly across all time points. In contrast, densities at locations $\bs_{500}$ and $\bs_{83014}$, which are farther apart, remain separated until the final time point, where the turbulence becomes more dispersed, and long-range dependence is observed.

\begin{figure}
    \centering
    \includegraphics[width=\linewidth]{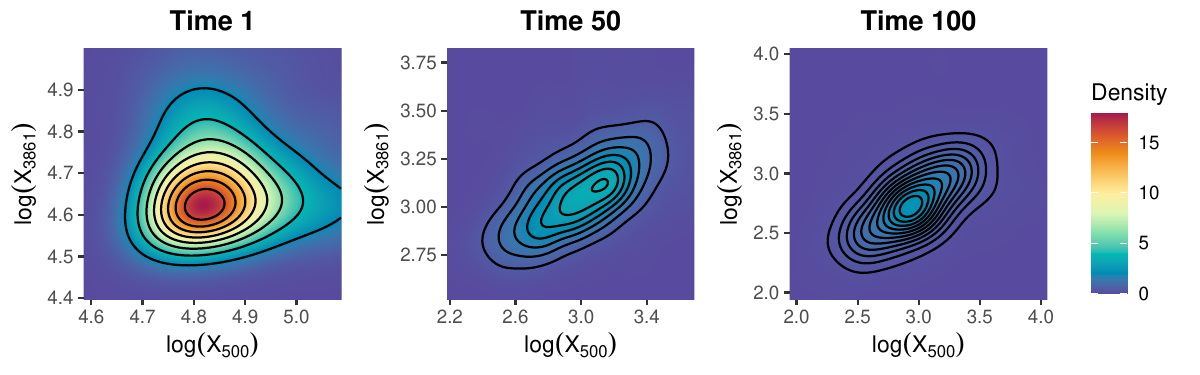}
    \includegraphics[width=\linewidth]{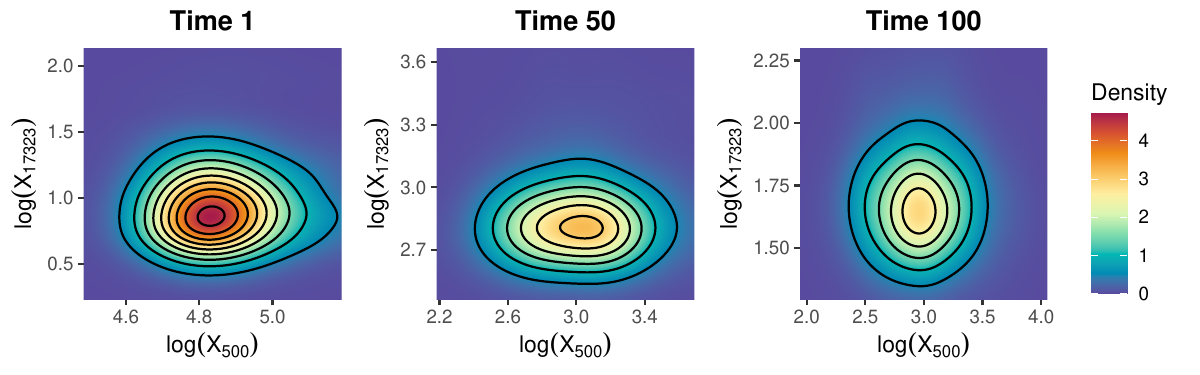}
    \includegraphics[width=\linewidth]{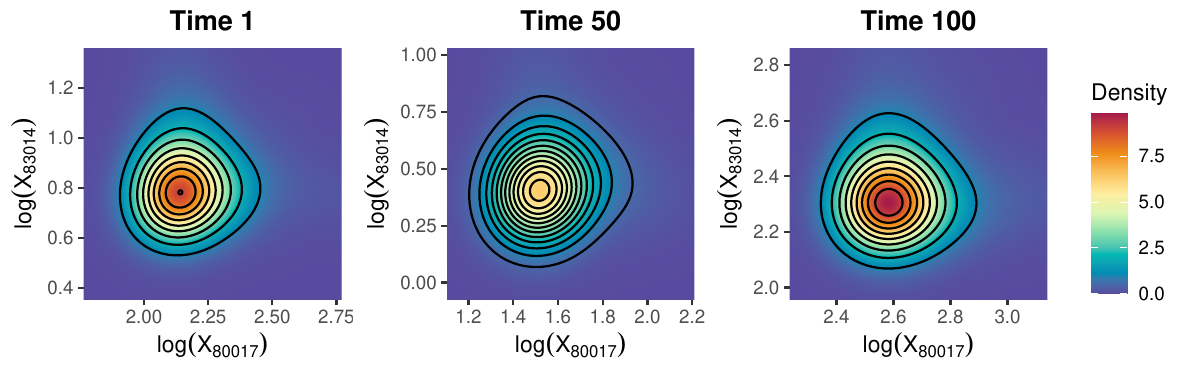}
    \caption{Bivariate posterior density of $\left(\log(X_{it}),\log(X_{jt})\right)^{\rm T}$ for $(i,j)^{\rm T} = (500, 3861)^{\rm T}$, $(500, 17323)^{\rm T}$, $(80017, 83014)^{\rm T}$ at $t=1,\; 50,\;100$. See Fig.~\ref{fig:ind_distr} for the changes in marginal distributions at these five locations.}
    \label{fig:kde2d}
\end{figure}

To analyze bivariate dependence among the five selected locations, we employ kernel density estimation (KDE) using the 1,000 emulated samples. We use a bivariate normal kernel with a bandwidth of $0.5$. Fig.~\ref{fig:kde2d} shows the joint distributions of $\left(\log(X_{it}),\log(X_{jt})\right)^{\rm T}$ for $i,\;j = 500$, $3861$, $17323$, $80017$, $83014$ at $t=1,\; 50,\;100$. The shift in marginal distributions, as observed in Fig.~\ref{fig:ind_distr}, is evident here as well. In the bivariate context, the differences in dependence become apparent: are relatively close, showing a higher positive correlation in their joint densities (first row of Fig.~\ref{fig:kde2d}), whereas while the other pairs of locations (middle and bottom rows) exhibit minimal correlation. Additionally, the movement of high joint density regions in the two-dimensional space over time is visible. As one can imagine, we can also examine changes in three-dimensional or higher-dimensional joint densities from the emulated samples.  

As seen in this section, accurate uncertainty quantification is crucial for understanding the range and likelihood of turbulence intensity at specific locations and times. This helps in assessing how sensitive the LES is to changes in input parameters and in identifying limitations or areas where the turbulent production mechanisms might not perform optimally. Particularly, with a more precise characterization of the joint tail, we can better evaluate the probability of extreme clusters, such as those involving the transport of hazardous or contaminant gases in the atmosphere. This is essential for improved risk management and more robust decision-making.

\section{Discussion}\label{sec:conclusion}

\par Wildland fire plumes is a complex turbulent flow with extreme turbulence due to  intense heat release, complex flow dynamics, and interactions between plumes and atmosphere. These turbulent plumes are highly energetic, intermittent, and exhibit chaotic behavior, creating significant challenges for modeling and predicting their behavior. Understanding the extreme regimes of turbulence within wildland fire plumes is crucial for predicting fire spread, fire behavior, and mitigating risks to both people and property. The turbulence within wildland fire plumes exhibits non-Gaussian, heavy-tailed behavior, especially when considering extreme events such as intense vortices, sudden bursts of energy, or large eddy structures. 

\par We compare the performance of POD and an VAE that can accommodate extremes (xVAE) for emulating a  turbulent buoyant plume. We use tail RMSE as a metric for cross-validation to optimize the number of basis functions or components used in these methods and compare their effectiveness in capturing extremal dependence and quantifying uncertainty.

The paper demonstrates that xVAE is a more effective and robust method for emulating this turbulent buoyant plume than low-order POD models, particularly for capturing extreme values
. We find that for both POD and xVAE, the number of basis functions significantly impacts performance. xVAE consistently outperforms POD with fewer bases in the tail RMSE, indicating better performance in capturing extremal dependence. POD, while effective for general distributions, struggles with extremal behavior due to oversmoothing. Unlike POD, xVAE provides sensible uncertainty quantification, offering valuable insights into the likelihood and range of extreme events, which is crucial for risk management and decision-making in turbulence analysis.

For future work, it will be interesting to conduct three-dimensional analysis to investigate how the scalar temperature field of buoyant turbulence changes in different directions of the three-dimensional space over time, instead of simply the center line. We can gain a more comprehensive understanding of the dynamic behavior of pure buoyant turbulent plume under study, potentially revealing patterns and correlations that are not apparent in two-dimensional analyses. 

We can also extend the xVAE framework to simultaneously emulate multiple variables, such as velocity, turbulent fluxes, and turbulent kinetic energy (TKE), to capture the complex interactions and dependencies between them. This requires implementing a multivariate POD or xVAE analysis, where the same basis functions are utilized across all dimensions of the data. By deriving a unified set of basis functions applicable to multiple variables, we can achieve a consistent representation of the underlying structures and patterns within the dataset. This approach enables us to learn extremal dependence over both space and variables, providing a deeper understanding of the multidimensional nature of the data and the interdependencies among variables.


We demonstrate in this study---for the first time---xVAE is a robust framework that can capture both the bulk and extreme features of turbulent flows. Traditional machine learning models, including neural networks, often struggle to generalize well in these scenarios due to the rarity and unpredictability of such extreme events, The methodology developed in this work is a novel framework for emulating extreme events in realistic turbulent flows such as atmospheric driven extreme events, oceanic turbulence, shock waves and detonation waves in the propulsion and compressible aerodynamic systems, and rocket plume emissions The study will provide an important direction for future studies that can transcend the current limits of computational needs.  Without the need for high-fidelity simulations that are computationally expensive. Using the xVAE, the neural networks can be trained more efficiently with high-quality synthetic data that can capture the extreme events in the flow.


\section{Methods}\label{sec:method_description}
\subsection{Proper orthogonal decomposition (POD)}
POD is commonly used in fluid dynamics to analyze turbulence. It is similar to principal component analysis (PCA) decomposition in statistics \cite[e.g., see the overview in][]{jolliffe2002principal}. The idea is to decompose a random field $\{X_t(\bs):\bs\in\mathcal{S}\subseteq\mathbb{R}^2\}$ into a set of empirical global basis functions $\{\phi_k(\bs):\bs\in\mathcal{S},\; k=1,\ldots, \infty\}$ modulated by random coefficients $\{a_{kt}:k=1,\ldots, \infty\}$ such that:
\begin{equation*}
    X_t(\bs)=\sum_{k=1}^\infty a_{kt} \phi_k(\bs),\;\bs\in\mathcal{S},
\end{equation*}
where $\mathcal{S}\in \mathbb{R}^2$ is the domain of interest.

\begin{figure}[!t]
    \centering
    \includegraphics[width=\linewidth]{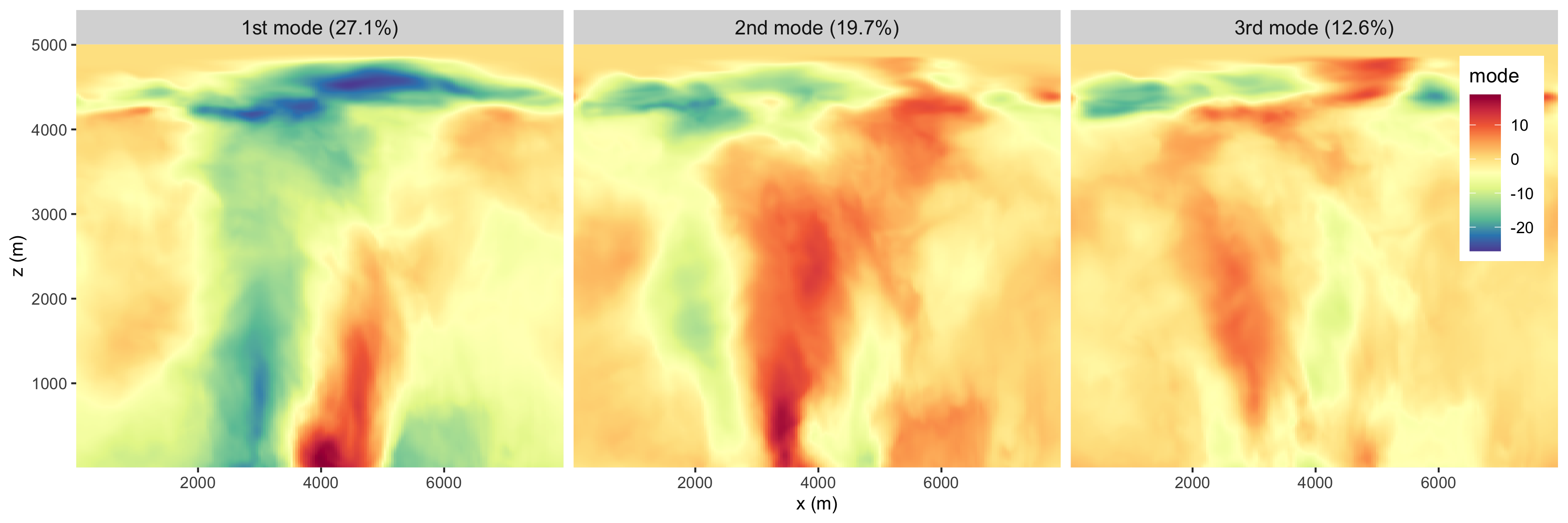}

    \includegraphics[width=\linewidth]{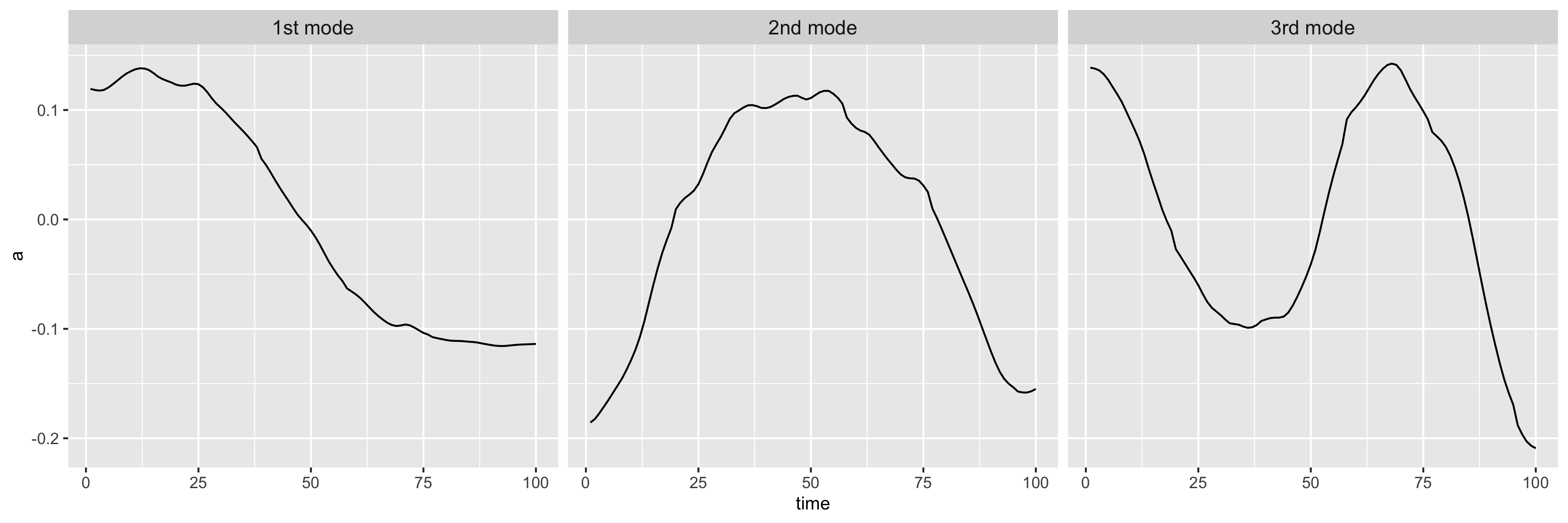}
    \caption{The first three modes from the POD of the turbulent buoyant plume  input and their corresponding coefficients $\{a_{kt}: k=1,\ldots,3, t=1,\ldots, n_t\}$.}
    \label{fig:pod-features}
\end{figure}

When the process $\{X_t(\bs)\}$ is observed/discretized at $n_s$ locations, $\bs_1, \ldots, \bs_{n_s}$, we call these observations at each time $t$ a snapshot of the field. Denote $X_{it} = X_t(\bs_i)$, $i=1,\ldots, n_s$, $\bX_t=(X_{1t},\ldots, X_{n_st})^{\rm T}$, and $\bX=(\bX_1,\ldots, \bX_{n_t})$. After centering each snapshot by their spatial averages,  we then perform singular value decomposition and obtain orthonormal matrices $\bW\in\mathbb{R}^{n_s\times n_s}$ and $\bV\in\mathbb{R}^{n_t\times n_t}$ such that $\bX=\bW\Lambda\bV^{\rm T}$. Here, $\Lambda\in\mathbb{R}^{n_s\times n_t}$ is a rectangular diagonal matrix with real diagonal elements, $\lambda_i$, corresponding to the eigenvalues of $\bX\bX^{\rm T}$ in decreasing order, i.e., $\lambda_1\geq \lambda_2\geq\cdots\geq \lambda_{n_s}$. Then the principle components or modes, $\bphi_1,\ldots, \bphi_{n_s}$, are simply the columns of $\bW\Lambda$ which are ranked based on their energy content (variance). It is common practice to use the first few modes to capture the most significant features of the data and to reduce dimensionality \cite[e.g.,][]{jolliffe2002principal}:
\begin{equation}\label{eqn:POD_reduce}
     \bX_t\approx\sum_{k=1}^{K} a_{kt} \bphi_k,
\end{equation}
where $K\ll n_t$ and $\ba_k = (a_{k1},\ldots, a_{kn_t} )^{\rm T}$ corresponds to the $k$th column of $\bV$, $1\leq k\leq K$. Fig.~\ref{fig:pod-features} shows visualizations of the first three modes and the corresponding coefficients across time of the plume simulated as described above. The first three and eight modes explain more than $59.40$\% and $91.14$\% of the variation in the input data, respectively.

As we can see from the bottom panels of Fig.~\ref{fig:pod-features}, the coefficients exhibit sinusoidal patterns with increasing frequencies as the eigenvalue decreases; see Fig.~\ref{fig:pod_coefs} in the Supplementary Material for the periodic pattern for the subsequent ten modes. Equation~\eqref{eqn:POD_reduce} allows for the reconstruction of the original data to approximate or visualize the data with reduced complexity, but as we will show in Section \ref{sec:results}, only summing a finite number of modes will oversmooth the process and inflate the extremal dependence. In addition, the POD does not incorporate probabilistic modeling and only provides a deterministic projection of data which does not account for uncertainty directly.

\subsection{VAE with spatial extremes}\label{sec:xVAE_description}
The turbulent buoyant plume is highly irregular and chaotic over space and volatile over time. Also, it is evaluated at a high-resolution grid, which makes the exact spatio-temporal modeling very challenging. To compound this issue, the turbulent buoyant plume  experience large clusters of extremes spatially that exhibit anisotropy. To characterize the long-range external dependence, we have to carefully model the tail of the joint distribution and also make such a model computationally efficient.

In \cite{zhang2023flexible}, a similar dimension-reduction technique to POD was applied so they could emulate the extreme values in Red Sea sea surface temperature in very high dimensions. Their model built upon the max-infinitely divisible (max-id) process, and extended it to allow for both short-range extremal (or asymptotic) dependence along with mid-range extremal (or asymptotic) independence, and long-range exact independence. 

Specifically, the scaled fluid density processes $\{X_t(\bs)\}$ can be modeled as follows:
\begin{equation}\label{eqn:model}
    X_t(\bs)=-\epsilon_t(\bs)Y_t(\bs),\;\bs\in\mathcal{S},
\end{equation}
where $\epsilon_t(\bs)$ is a noise process with independent Fr\'{e}chet$(0,\tau,1/\alpha_0)$ marginal distributions; that is, $\Pr\{\epsilon_t(\bs)\leq x\}=\exp\{-(x/\tau)^{1/\alpha_0}\}$, where $x>0$, $\tau>0$ and $\alpha_0>0$. Then, $Y_t(\bs)$ is constructed using a low-rank representation similar to equation~\eqref{eqn:POD_reduce}: 
\begin{equation}\label{eqn:low_rank_representation}
    Y_t(\bs)=\left\{\sum_{k=1}^K \omega_k(\bs) Z_{kt}\right\}^{\alpha_0},
\end{equation}
where $\alpha\in (0,1)$, $\{\omega_k(\bs): \bs\in\mathcal{S},k=1,\ldots, K\}$ are a set of basis functions, and $\{Z_{kt}:k=1,\ldots,K\}$ are latent variables that have independent exponentially-tilted positive-stable (PS) distributions, denoted as $Z_{kt}\stackrel{\text{ind}}{\sim} \mathrm{expPS}(\alpha,\theta_k),\; k=1,\ldots, K$, $t=1,\ldots, n_t$. Here, the parameter $\alpha\in (0,1)$ determines the rate at which the power-law tail of $\mathrm{expPS}(\alpha,0)$ tapers off, and the tilting parameter $\theta_k\geq 0$ determines the extent of tilting, with larger values of $\theta_k$ leading to lighter-tailed $\mathrm{expPS}(\alpha,\theta_k)$. See Section A.1 in \cite{zhang2023flexible} for more specific forms of a $\mathrm{expPS}(\alpha,\theta_k)$ distribution.

The construct in equations~\eqref{eqn:model}--\eqref{eqn:low_rank_representation} achieves max-infinite divisibility through the Fréchet noise, whose heavy-tailed behavior aligns with the power-law characteristics of the expPS variables. Max-infinite divisibility is useful for spatial extremes modeling because it enables the representation of realistic spatial patterns of extreme events while maintaining mathematical tractability. By integrating expPS distributions, the model benefits from a clean Laplace transform and localized control of tail dependence through tuning the values of $\theta_k$'s, enhancing flexibility and precision in capturing extreme behaviors. These features make the framework highly effective for applications such as extreme weather modeling and risk assessment.

More importantly, the hierarchical construct in equations~\eqref{eqn:model}--\eqref{eqn:low_rank_representation} also allows the embedding of the latent variables $Z_{kt}, k=1,\ldots,\ K$ in encoded space of the VAE; that is, we compress/encode the spatio-temporal inputs from the marginally transformed turbulent buoyant plume  to the much smaller $K$-dimensional latent space, while keeping the distributional assumptions in equations~\eqref{eqn:model}--\eqref{eqn:low_rank_representation} when decompressing/decoding the latent variables back to the initial space. Both the encoder and decoder are multilayer perceptron (MLP) neural networks whose weights are updated iteratively to minimize the differentiable Evidence Lower Bound (ELBO). Fig.~\ref{fig:vae_diagram} illustrates a schematic overview of the xVAE framework, while Section~\ref{sec:xvae_details} in the Supplementary Material provides a comprehensive explanation of the model's architecture and implementation details.

Once the xVAE is trained and the parameters for the neural networks are optimized, posterior simulation of new latent variables $\bZ_t$ for the observations $\bX_t$ at time $t$ can be performed efficiently using the encoder. Synthetic data $\bX_t^*$ can then be generated rapidly by passing these latent variables through the decoder and sampling from the model $\bX_t\sim p_{\bphi_d}(\bx_t\mid\bz_t)$ specified by equations~\eqref{eqn:model} and \eqref{eqn:low_rank_representation}. 

\subsection{Non-negative matrix factorization}
In the Red Sea sea surface temperature study, \cite{zhang2023flexible} utilized locally compact Wendland basis functions in place of $\{\omega_k(\bs): \bs\in\mathcal{S},k=1,\ldots, K\}$ in equation~\eqref{eqn:low_rank_representation}. But for the turbulent buoyant plume, the local basis functions will dampen the irregularity of the local extremes. Global data-driven basis functions are better suited for this chaotic turbulence data.

POD offers a good candidate set of basis functions $\{\omega_k(\bs): \bs\in\mathcal{S},k=1,\ldots, K\}$ for xVAE, but a good representation of $\{X_t(\bs)\}$ at one time $t$ using POD bases will require some of the random coefficients $\{Z_{kt}:k=1,\ldots, K\}$ to be negative. However, the low-rank decomposition in equation~\eqref{eqn:low_rank_representation} uses random coefficients that are always positive. To address this, we use Non-Negative Matrix Factorization (NMF) to find bases that allow us to project $\{X_t(\bs)\}$ onto the latent space with positive coefficients.

NMF is a dimension reduction technique widely used in many applications such as image processing, text mining, and spectral data analysis \cite{paatero1994positive}. NMF imposes a nonnegativity constraint for matrix factorization, which facilitates learning a parts-based representation of the data \cite{lee2000algorithms}. NMF is also better at extracting hidden features that are more sparse and easily interpretable while keeping the structure of the original data intact from a non-negative matrix \cite{gillis2014and}.

\begin{figure}[!t]
    \centering
    \includegraphics[width=\linewidth]{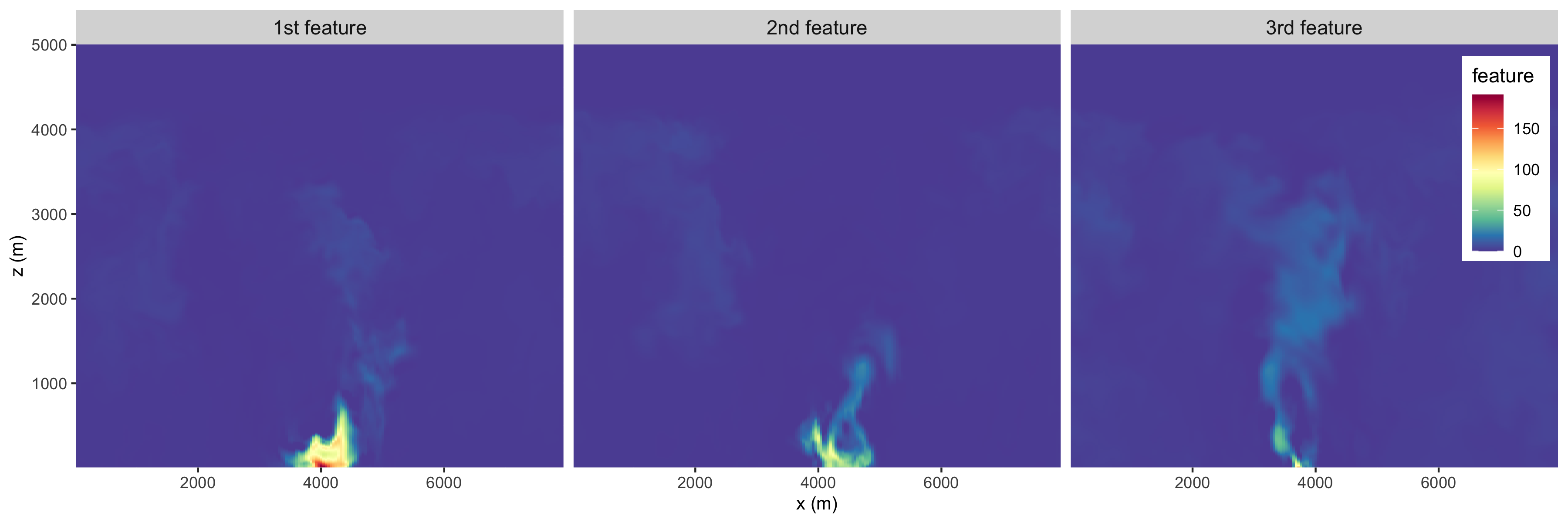}

    \includegraphics[width=\linewidth]{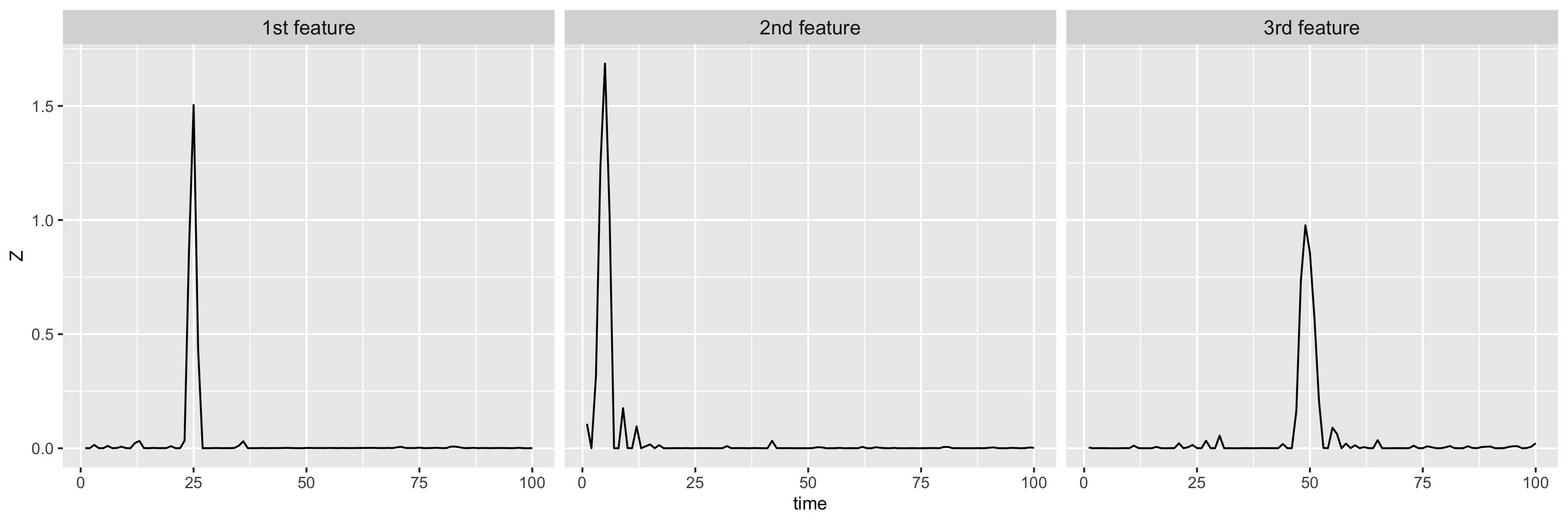}
    \caption{The first three features from the non-negative matrix factorization of the turbulent buoyant plume  input and their corresponding initial coefficients over time $\{Z^0_{kt}: k=1,\ldots,3, t=1,\ldots, n_t\}$, i.e., the first three rows of $\bZ^0$. These coefficients are mostly near zero and exhibit a single peak, suggesting minimal autocorrelation and evidence of heavy-tailed behavior.}
    \label{fig:nmf-features}
\end{figure}

Again, we discretize the spatial processes $n_s$ locations $\{\bs_1,\ldots, \bs_{n_s}\}$. We can use NMF to extract a $n_s\times K$ feature matrix $\bOmega = \{\omega_k(\bs_i): k=1,\ldots,K, i=1,\ldots, n_s\}\equiv (\bomega_1, \ldots, \bomega_K)$ from the original data matrix $\bX=\{X_t(\bs_i): t=1,\ldots,n_t, i=1,\ldots, n_s\}$:
\begin{equation*}
    \bX \approx -\bOmega \bZ^0,  
\end{equation*}
where all three matrices have no negative elements and $K\ll n_t$; see Fig.~\ref{fig:nmf-features} for the visualization of the first three features. The feature matrix and the component matrix are derived via minimizing the Frobenius distance $||\bX + \bOmega \bZ^0||_F$ under the non-negativity constraint, where $||\cdot||_F$ is obtained via taking the square root of the sum of the squares of all matrix entries. This minimization is implemented using the Coordinate Descent solver in the \texttt{sklearn.decomposition} module in \texttt{python} \cite{pedregosa2011scikit}. We will use the column vectors of $\bOmega$ as bases in the modeling of latent process at the observed locations and times. i.e., $\bY=\{Y_t(\bs_i): t=1,\ldots,n_t, i=1,\ldots, n_s\}$. 

Using the features from NMF is advantageous for applying our max-infinitely divisible model because both the bases $(\bomega_1, \ldots, \bomega_K)$ and initial coefficients $\bZ^0$ are non-negative. Moreover, the basis vectors only need to be calculated once before the spatial extremes analysis. In contrast, the bases $ (\bphi_1, \ldots, \bphi_K)$ obtained from POD  will lead to basis coefficients that are heavy-tailed on both sides of $\mathbb{R}$. Modeling one-sided heavy tails and their dependence structure is easier and more tractable. It should noted that the matrix $\bZ^0$ is just an initial approximate of the random coefficients $\{Z_{kt}: k=1,\ldots,K, t=1,\ldots, n_t\}$. However, $\bZ^0$ informs us of the general scale of the coefficients. The bottom panels in Fig.~\ref{fig:nmf-features} indicate that modeling $\{Z_{kt}:t=1,\ldots, n_t\}$ for a fixed $k$ with independent exponentially-tilted PS variables is reasonable, as each row of $\bZ^0$ shows heavy-tailedness and minimal autocorrelation. Additionally, the clustering property of NMF, where features are prominent during consecutive time periods, and the sparse nature of the feature matrix $\bOmega$, aid in identifying meaningful patterns and also lead to more efficient storage and computation, especially given the large number of spatial locations $n_s$.

\section*{Competing interests}
We declare that none of the authors have competing financial or non-financial interests as defined by Nature Portfolio.

\section*{Inclusion \& Ethics}
No human participants or animal subjects were involved in this study, and therefore no specific ethical approval was required. The authors confirm that all work was conducted in accordance with relevant standards and guidelines, and that the study design and execution did not involve any activities requiring institutional review board oversight. We are committed to fostering an inclusive and equitable research environment, and all authors strived to ensure that research practices were fair, transparent, and considerate of diverse perspectives throughout the study.

\section*{Data availability}
All datasets, code, and post-processing routines used in this study are publicly available at \href{https://github.com/CFD-UTSA/WRF-bLES-PyPlume}{github.com/CFD-UTSA/WRF-bLES-PyPlume}. There are no restrictions on data availability.

\section*{Code availability}
All custom code and algorithms used in this study are freely available at the GitHub repository(\href{https://github.com/likun-stat/plume_analysis}{github.com/likun-stat/plume\_analysis}) under the MIT license. Detailed instructions for downloading, installing, and running the code are provided in the repository’s README file.

\addcontentsline{toc}{section}{References}
\printbibliography

@book{jolliffe2002principal,
  title={Principal component analysis for special types of data},
  author={Jolliffe, Ian T},
  year={2002},
  publisher={Springer}
}

@article{bhaganagar2020numerical,
  title={Numerical investigation of starting turbulent buoyant plumes released in neutral atmosphere},
  author={Bhaganagar, Kiran and Bhimireddy, Sudheer R},
  journal={Journal of Fluid Mechanics},
  volume={900},
  pages={A32},
  year={2020},
  publisher={Cambridge University Press}
}

@article{britter1989atmospheric,
  title={Atmospheric dispersion of dense gases},
  author={Britter, Rex E},
  journal={Annual review of fluid mechanics},
  volume={21},
  number={1},
  pages={317--344},
  year={1989}
}

@article{bhaganagar2020local,
  title={Local atmospheric factors that enhance air-borne dispersion of coronavirus-High-fidelity numerical simulation of COVID19 case study in real-time},
  author={Bhaganagar, Kiran and Bhimireddy, Sudheer},
  journal={Environmental research},
  pages={110170},
  year={2020},
  publisher={Elsevier}
}

@article{bhaganagar2017assessment,
  title={Assessment of the plume dispersion due to chemical attack on April 4, 2017, in Syria},
  author={Bhaganagar, Kiran and Bhimireddy, Sudheer R},
  journal={Natural hazards},
  volume={88},
  number={3},
  pages={1893--1901},
  year={2017},
  publisher={Springer}
}

@article{chen2023energetics,
  title={Energetics of buoyancy-generated turbulent flows with active scalar: pure buoyant plume},
  author={Chen, Chang Hsin and Bhaganagar, Kiran},
  journal={Journal of Fluid Mechanics},
  volume={954},
  pages={A23},
  year={2023},
  publisher={Cambridge University Press}
}

@article{chen2021new,
  title={New findings in vorticity dynamics of turbulent buoyant plumes},
  author={Chen, Chang Hsin and Bhaganagar, Kiran},
  journal={Physics of Fluids},
  volume={33},
  number={11},
  year={2021},
  publisher={AIP Publishing}
}

@article{Lumley93,
  title={The proper orthogonal decomposition in the analysis of turbulent flows},
  author={Berkooz, Gal and Holmes, Philip and Lumley, John L},
  journal={Annual review of fluid mechanics},
  volume={25},
  number={1},
  pages={539--575},
  year={1993},
  publisher={Annual Reviews}
}

@article{cammilleri2013pod,
  title={POD-spectral decomposition for fluid flow analysis and model reduction},
  author={Cammilleri, Ada and Gu{\'e}niat, Florimond and Carlier, Johan and Pastur, Luc and M{\'e}min, Etienne and Lusseyran, Fran{\c{c}}ois and Artana, Guillermo},
  journal={Theoretical and Computational Fluid Dynamics},
  volume={27},
  pages={787--815},
  year={2013},
  publisher={Springer}
}

@article{hijazi2020data,
  title={Data-driven POD-Galerkin reduced order model for turbulent flows},
  author={Hijazi, Saddam and Stabile, Giovanni and Mola, Andrea and Rozza, Gianluigi},
  journal={Journal of Computational Physics},
  volume={416},
  pages={109513},
  year={2020},
  publisher={Elsevier}
}

@article{schubert2022towards,
  title={Towards robust data-driven reduced-order modelling for turbulent flows: application to vortex-induced vibrations},
  author={Schubert, Yannick and Sieber, Moritz and Oberleithner, Kilian and Martinuzzi, Robert},
  journal={Theoretical and Computational Fluid Dynamics},
  volume={36},
  number={3},
  pages={517--543},
  year={2022},
  publisher={Springer}
}

@article{schmid2022dynamic,
  title={Dynamic mode decomposition and its variants},
  author={Schmid, Peter J},
  journal={Annual Review of Fluid Mechanics},
  volume={54},
  number={1},
  pages={225--254},
  year={2022},
  publisher={Annual Reviews}
}

@article{bhaganagar2021MWR,
  title={Implementing a new formulation in WRF-LES for Buoyant Plume Simulations: bPlume-WRF-LES model},
  author={Bhimireddy, Sudheer R and Bhaganagar, Kiran},
  journal={Monthly Weather Review},
  volume={149},
  number={7},
  pages={2299-2319},
  year={2021},
}

@article{baddoo2023physics,
  title={Physics-informed dynamic mode decomposition},
  author={Baddoo, Peter J and Herrmann, Benjamin and McKeon, Beverley J and Nathan Kutz, J and Brunton, Steven L},
  journal={Proceedings of the Royal Society A},
  volume={479},
  number={2271},
  pages={20220576},
  year={2023},
  publisher={The Royal Society}
}

@article{fukami2020convolutional,
  title={Convolutional neural network based hierarchical autoencoder for nonlinear mode decomposition of fluid field data},
  author={Fukami, Kai and Nakamura, Taichi and Fukagata, Koji},
  journal={Physics of Fluids},
  volume={32},
  number={9},
  year={2020},
  publisher={AIP Publishing}
}

@inproceedings{momenifar2022physics,
  title={A physics-informed vector quantized autoencoder for data compression of turbulent flow},
  author={Momenifar, Mohammadreza and Diao, Enmao and Tarokh, Vahid and Bragg, Andrew D},
  booktitle={2022 Data Compression Conference (DCC)},
  pages={01--10},
  year={2022},
  organization={IEEE}
}

@article{glaws2020deep,
  title={Deep learning for in situ data compression of large turbulent flow simulations},
  author={Glaws, Andrew and King, Ryan and Sprague, Michael},
  journal={Physical Review Fluids},
  volume={5},
  number={11},
  pages={114602},
  year={2020},
  publisher={APS}
}

@article{solera2024beta,
  title={$\beta$-Variational autoencoders and transformers for reduced-order modelling of fluid flows},
  author={Solera-Rico, Alberto and Sanmiguel Vila, Carlos and G{\'o}mez-L{\'o}pez, Miguel and Wang, Yuning and Almashjary, Abdulrahman and Dawson, Scott TM and Vinuesa, Ricardo},
  journal={Nature Communications},
  volume={15},
  number={1},
  pages={1361},
  year={2024},
  publisher={Nature Publishing Group UK London}
}

@article{linot2023dynamics,
  title={Dynamics of a data-driven low-dimensional model of turbulent minimal Couette flow},
  author={Linot, Alec J and Graham, Michael D},
  journal={Journal of Fluid Mechanics},
  volume={973},
  pages={A42},
  year={2023},
  publisher={Cambridge University Press}
}

@article{kolmogorov1991local,
  title={The local structure of turbulence in incompressible viscous fluid for very large Reynolds numbers},
  author={Kolmogorov, Andrei Nikolaevich},
  journal={Proceedings of the Royal Society of London. Series A: Mathematical and Physical Sciences},
  volume={434},
  number={1890},
  pages={9--13},
  year={1991},
  publisher={The Royal Society London}
}

@article{moffatt2021extreme,
  title={Extreme events in turbulent flow},
  author={Moffatt, HK},
  journal={Journal of Fluid Mechanics},
  volume={914},
  pages={F1},
  year={2021},
  publisher={Cambridge University Press}
}

@article{l2001outliers,
  title={Outliers, extreme events, and multiscaling},
  author={L’vov, Victor S and Pomyalov, Anna and Procaccia, Itamar},
  journal={Physical Review E},
  volume={63},
  number={5},
  pages={056118},
  year={2001},
  publisher={APS}
}

@book{frisch1995turbulence,
  title={Turbulence: the legacy of AN Kolmogorov},
  author={Frisch, Uriel},
  year={1995},
  publisher={Cambridge university press}
}

@article{gotoh2022transition,
  title={Transition of fluctuations from Gaussian state to turbulent state},
  author={Gotoh, Toshiyuki and Yang, Jingyuan},
  journal={Philosophical Transactions of the Royal Society A},
  volume={380},
  number={2218},
  pages={20210097},
  year={2022},
  publisher={The Royal Society}
}

@article{driscoll2020premixed,
  title={Premixed flames subjected to extreme turbulence: Some questions and recent answers},
  author={Driscoll, James F and Chen, Jacqueline H and Skiba, Aaron W and Carter, Campbell D and Hawkes, Evatt R and Wang, Haiou},
  journal={Progress in Energy and Combustion Science},
  volume={76},
  pages={100802},
  year={2020},
  publisher={Elsevier}
}

@article{hassanaly2021classification,
  title={Classification and computation of extreme events in turbulent combustion},
  author={Hassanaly, Malik and Raman, Venkat},
  journal={Progress in Energy and Combustion Science},
  volume={87},
  pages={100955},
  year={2021},
  publisher={Elsevier}
}

@article{sapsis2021statistics,
  title={Statistics of extreme events in fluid flows and waves},
  author={Sapsis, Themistoklis P},
  journal={Annual Review of Fluid Mechanics},
  volume={53},
  number={1},
  pages={85--111},
  year={2021},
  publisher={Annual Reviews}
}

@article{farazmand2017variational,
  title={A variational approach to probing extreme events in turbulent dynamical systems},
  author={Farazmand, Mohammad and Sapsis, Themistoklis P},
  journal={Science advances},
  volume={3},
  number={9},
  pages={e1701533},
  year={2017},
  publisher={American Association for the Advancement of Science}
}

@article{duraisamy2019turbulence,
  title={Turbulence modeling in the age of data},
  author={Duraisamy, Karthik and Iaccarino, Gianluca and Xiao, Heng},
  journal={Annual review of fluid mechanics},
  volume={51},
  number={1},
  pages={357--377},
  year={2019},
  publisher={Annual Reviews}
}

@article{zare2020stochastic,
  title={Stochastic dynamical modeling of turbulent flows},
  author={Zare, Armin and Georgiou, Tryphon T and Jovanovi{\'c}, Mihailo R},
  journal={Annual Review of Control, Robotics, and Autonomous Systems},
  volume={3},
  number={1},
  pages={195--219},
  year={2020},
  publisher={Annual Reviews}
}

@article{sharman2016nature,
  title={Nature of aviation turbulence},
  author={Sharman, Robert},
  journal={Aviation turbulence: Processes, detection, prediction},
  pages={3--30},
  year={2016},
  publisher={Springer}
}

@article{beck2021perspective,
  title={A perspective on machine learning methods in turbulence modeling},
  author={Beck, Andrea and Kurz, Marius},
  journal={GAMM-Mitteilungen},
  volume={44},
  number={1},
  pages={e202100002},
  year={2021},
  publisher={Wiley Online Library}
}

@article{fukami2023grasping,
  title={Grasping extreme aerodynamics on a low-dimensional manifold},
  author={Fukami, Kai and Taira, Kunihiko},
  journal={Nature Communications},
  volume={14},
  number={1},
  pages={6480},
  year={2023},
  publisher={Nature Publishing Group UK London}
}

@article{buaria2019extreme,
  title={Extreme velocity gradients in turbulent flows},
  author={Buaria, Dhawal and Pumir, Alain and Bodenschatz, Eberhard and Yeung, Pui-Keun},
  journal={New Journal of Physics},
  volume={21},
  number={4},
  pages={043004},
  year={2019},
  publisher={IOP Publishing}
}

@article{goldenfeld2010extreme,
  title={Extreme fluctuations and the finite lifetime of the turbulent state},
  author={Goldenfeld, Nigel and Guttenberg, Nicholas and Gioia, Gustavo},
  journal={Physical Review E—Statistical, Nonlinear, and Soft Matter Physics},
  volume={81},
  number={3},
  pages={035304},
  year={2010},
  publisher={APS}
}

@article{yeung2015extreme,
  title={Extreme events in computational turbulence},
  author={Yeung, PK and Zhai, XM and Sreenivasan, Katepalli R},
  journal={Proceedings of the National Academy of Sciences},
  volume={112},
  number={41},
  pages={12633--12638},
  year={2015},
  publisher={National Acad Sciences}
}

@article{xia2023hierarchical,
  title={A hierarchical autoencoder and temporal convolutional neural network reduced-order model for the turbulent wake of a three-dimensional bluff body},
  author={Xia, Chao and Wang, Mengjia and Fan, Yajun and Yang, Zhigang and Du, Xuzhi},
  journal={Physics of Fluids},
  volume={35},
  number={2},
  year={2023},
  publisher={AIP Publishing}
}

@article{skamarock2008description,
  title={A description of the advanced research WRF version 3},
  author={Skamarock, William C and Klemp, Joseph B and Dudhia, Jimy and Gill, David O and Barker, Dale M and Duda, Michael G and Huang, Xiang-Yu and Wang, Wei and Powers, Jordan G and others},
  journal={NCAR technical note},
  volume={475},
  pages={113},
  year={2008},
  publisher={NCAR Boulder, CO, USA}
}

@article{zhang2023flexible,
  title={Flexible and efficient spatial extremes emulation via variational autoencoders},
  author={Zhang, Likun and Ma, Xiaoyu and Wikle, Christopher K and Huser, Rapha{\"e}l},
  journal={Preprint at https://arxiv.org/abs/2307.08079},
  year={2024}
}

@article{lee2000algorithms,
  title={Algorithms for non-negative matrix factorization},
  author={Lee, Daniel and Seung, H Sebastian},
  journal={Advances in neural information processing systems},
  volume={13},
  year={2000}
}

@article{gillis2014and,
  title={The why and how of nonnegative matrix factorization},
  author={Gillis, Nicolas},
  journal={Regularization, optimization, kernels, and support vector machines},
  volume={12},
  number={257},
  pages={257--291},
  year={2014},
  publisher={Chapman \& Hall/CRC Boca Raton}
}

@article{paatero1994positive,
  title={Positive matrix factorization: A non-negative factor model with optimal utilization of error estimates of data values},
  author={Paatero, Pentti and Tapper, Unto},
  journal={Environmetrics},
  volume={5},
  number={2},
  pages={111--126},
  year={1994},
  publisher={Wiley Online Library}
}

@article{kingma2013auto,
  title={Auto-encoding variational {B}ayes},
  author={Kingma, Diederik P and Welling, Max},
  journal={Preprint at https://arxiv.org/abs/1312.6114},
  year={2013}
}

@article{pedregosa2011scikit,
  title={Scikit-learn: Machine learning in Python},
  author={Pedregosa, Fabian and Varoquaux, Ga{\"e}l and Gramfort, Alexandre and Michel, Vincent and Thirion, Bertrand and Grisel, Olivier and Blondel, Mathieu and Prettenhofer, Peter and Weiss, Ron and Dubourg, Vincent and others},
  journal={the Journal of machine Learning research},
  volume={12},
  pages={2825--2830},
  year={2011},
  publisher={JMLR. org}
}

@article{davison2015statistics,
  title={Statistics of extremes},
  author={Davison, Anthony C. and Huser, Rapha\"el},
  journal={Annual Review of Statistics and its Application},
  volume={2},
  year={2015},
  pages = {203--235}
}

@article{huser2022advances,
  title={Advances in statistical modeling of spatial extremes},
  author={Huser, Rapha{\"e}l and Wadsworth, Jennifer L},
  journal={Wiley Interdisciplinary Reviews: Computational Statistics},
  volume={14},
  number={1},
  pages={e1537},
  year={2022},
  publisher={Wiley Online Library}
}

@article{stolovitzky1993scaling,
  title={Scaling of structure functions},
  author={Stolovitzky, G and Sreenivasan, KR},
  journal={Physical Review E},
  volume={48},
  number={1},
  pages={R33},
  year={1993},
  publisher={APS}
}

@article{andrew2025amortized,
  title={Neural Methods for Amortised Parameter Inference},
  author={Zammit-Mangion, Andrew and Sainsbury-Dale, Matthew and Huser, Rapha{\"e}l},
  journal={Annual Reviews of Statistics and Its Application},
  year={2025},
  note={to appear}
}

@inproceedings{zhang2024capturing,
  title={Capturing Extreme Events in Turbulence using an Extreme Variational Autoencoder},
  author={Zhang, Likun and Wikle, Christopher and Bhaganagar, Kiran},
  booktitle={NeurIPS 2024 Workshop on Bayesian Decision-making and Uncertainty},
    year={2024}
}
\appendix

\section{WRF-LES model details}\label{appen:wrf-les}
The WRF-LES model solves the compressible Euler equations in their conservative flux formulation. These governing equations include the conservation of mass \eqref{MassCons_LES}, momentum \eqref{UMomentumCons_LES}, and energy \eqref{EnergyCons_LES}, which collectively describe the dynamics of the flow.
\begin{align}
    \partial_{t}\rho + \partial_i({\rho u_i}) &= 0,
  \label{MassCons_LES}\\
  \partial_{t}(\rho u_i) + \partial_j( {\rho u_i} u_j) +\partial_{i}p +\delta_{i3}\rho g &= \partial_j (\rho K_j \partial_j u_i)+\partial_i(\rho K_j\partial_j u_j),
  \label{UMomentumCons_LES}\\
  \partial_{t}(\rho \theta) + \partial_i({\rho u_i}\theta) &= \partial_i(\rho K_{i}^{h}\partial_i\theta)-\delta_{i3}\rho g H_o,
  \label{EnergyCons_LES}
\end{align}
where $u_i$ represents the velocities in $x,y$ and $z$ directions, $p$ is the pressure, $\theta$ is the potential temperature, $\rho$ is the density, $g$ is acceleration due to gravity and $K_{i}$ is the horizontal ($i=1,2$) and vertical ($i=3$) eddy viscosity determined by 1.5-order TKE closure scheme. In addition, $l_{i}$ represents the mixing length in horizontal ($i=1,2$) and vertical ($i=3$), and $e$ is subgrid TKE, which is predicted by 
\begin{equation*}
\partial_{t}(\rho e) + \partial_i(\rho u_i e) = \rho K_i[\partial_i u_j+\partial_j u_i]^2-\rho K_{3}N^2 - \rho\frac{Ce^{3/2}}{l},
\label{TKE_Equ}
\end{equation*}
where $N$ is the Brunt-V\"ais\"al\"a frequency calculated as $N^2 = (g/\theta) \partial_{z}\theta$. In addition, 
\begin{equation*}
    C = 1.9 C_k +\frac{\max\{0,0.93-1.9C_k\}l}{\Delta{s}},
\end{equation*}
with $\Delta{s}=\sqrt[3] {\Delta{x}\Delta{y}\Delta{z}}$, and $l=\min\{\Delta{s},0.76\sqrt{e}/N\}$, and $\Delta{x},\Delta{y},\Delta{z}$ represent the grid resolution in horizontal and vertical, respectively. Based on the potential temperature gradient, the horizontal and vertical mixing lengths are estimated. For $N^2>0$, $l_{1,2}=l_{3}=l$ and for $N^2\leq 0$, $l_{1,2}=l_{3}=\Delta{s}$. Finally, $K_{i}^h$ in equation (\ref{EnergyCons_LES}) represents the horizontal ($i=1,2$) and vertical ($i=3$) eddy diffusivity of heat estimated by multiplying the earlier predicted $K_{i}$ with $1+2(l_{i}/\Delta{s})$.

\section{xVAE details}\label{sec:xvae_details}
The xVAE framework integrates a VAE design with the max-id model to handle extreme spatial data. For $t=1,\ldots, n_t$, the encoder in the xVAE maps each observed spatial replicate $\bX_t$ to distributions over the latent space, $q_{\bphi_e}(\bz_t\mid \bx_t)$, using a standard reparameterization trick with an auxiliary variable $\boldsymbol{\eta}_t$, i.e.,
\begin{equation}\label{eqn:encoder_form}
 \begin{split}
        (\bmu_t^{\rm T},\log \bzeta_t^{\rm T})^{\rm T} &= \mathrm{EncoderNeuralNet}_{\bphi_e}(\bX_t),\\
    \eta_{kt}&\stackrel{\text{i.i.d.}}{\sim} \mathrm{Normal}(0,1),\\
    \bZ_t&=\bmu_t + \bzeta_t \odot \boldsymbol{\eta}_t,
 \end{split}
\end{equation}
in which $\bphi_e$ denotes the weights and biases of the encoder neural networks, $\boldsymbol{\eta}_t=\{\eta_{kt}:k=1,\ldots, K\}$ and $\odot$ is the elementwise product. 

Then the decoder learns the dependence parameters $(\alpha_t,\btheta_t^{\rm T})^{\rm T}$ and reconstructs data from the generated $\bZ_t$, $t=1, \ldots, n_t$. Specifically,
\begin{equation}\label{eqn:decoder_form}
    (\alpha_t, \btheta_t^{\rm T})^{\rm T} = \mathrm{DecoderNeuralNet}_{\bphi_d}(\bZ_t),
\end{equation}
where $\bphi_d$ includes the weights and biases of the decoder neural networks and also the Fr\'{e}chet noise parameters $(\alpha_0,\tau)$. To reconstruct the data from $\bZ_t$, we sample independent Fr\'{e}chet-distributed noises and combine it with latent variables $\bZ_t$ according to the model's governing equations Eqs.~\eqref{eqn:model}--\eqref{eqn:low_rank_representation}.

To retain the maximum amount of information when encoding while minimizing the reconstruction error when decoding, we maximize the objective function---evidence lower bound (ELBO), which balances the log-likelihood and the Kullback--Leibler (KL) divergence between the approximate and true posteriors. It can be written as
\begin{equation*}
\begin{split}
\mathcal{L}_{\bphi_e,\bphi_d}(\bx_1, \ldots, \bx_t) &= \sum_{t=1}^{n_t}\log p_{\bphi_d}(\bx_t)- \sum_{t=1}^{n_t}D_{KL}\left\{q_{\bphi_e}(\bz_t\mid\bx_t)\;||\;p_{\bphi_d}(\bz_t\mid \bx_t)\right\}\\
&=\mathbb{E}_{ q_{\bphi_e}(\bz_t\mid\bx_t)}\left\{\log\frac{p_{\bphi_d}(\bx_t,\bz_t)}{q_{\bphi_e}(\bz_t\mid\bx_t)}\right\},
\end{split}
\end{equation*} 
in which $q_{\bphi_e}(\bz_t\mid\bx_t)$ is defined by the reparameterization trick in Eq.~\eqref{eqn:encoder_form}, and $p_{\bphi_d}(\bx_t,\bz_t)=p_{\bphi_d}(\bx_t\mid \bz_t)p_{\bphi_d}(\bz_t)$. Further, $p_{\bphi_d}(\bx_t\mid \bz_t)$ is derived from differentiating the conditional distribution function
 derived from Eqs.~\eqref{eqn:model}--\eqref{eqn:low_rank_representation}:
\begin{equation*}\label{eqn:lik_form}
     \Pr(\bX_t\leq \bx_t\mid \bz_t,\bphi_d)= \exp\left\{-\sumN \left({\tau}/{x_{jt}}\right)^{{1}/{\alpha_0}}\sumK\omega_{kj}^{{1}/{\alpha_t}}z_{kt}\right\},
\end{equation*}
Denoting the density function of the $\mathrm{expPS}(\alpha,\theta_{k})$ distribution by $h(z;\alpha,\theta_{k})$, the prior on $\bz_t$ can be written as
\begin{equation*}\label{eqn:prior_form}
    p_{\bphi_d}(\bz_t) = \prodK h(z_{kt};\alpha_t,\theta_{kt}).
\end{equation*}
Crucially, the reparameterization trick enables fast computation of Monte Carlo estimates of $\nabla_{\bphi_e}\mathcal{L}_{\bphi_e, \bphi_d}$, the gradient of the ELBO with respect to $\bphi_e$. In contrast, $\nabla_{\bphi_d}\mathcal{L}_{\bphi_e, \bphi_d}$ is more straightforward because the ELBO’s expectation is taken only with respect to the distribution $q_{\bphi_e}(\bz_t\mid\bx_t)$.

Therefore, the xVAE allows for joint optimization over all parameters ($\bphi_e$ and $\bphi_d$) using stochastic gradient descent. Once the optimum parameters  are determined, we can sample large number of $Z_t$ from the encoder Eq.~\eqref{eqn:encoder_form}, and reconstruct copies of the original field $\bX_t$ that have similar properties, especially in the joint tail. The generative capacity of the xVAE allows for rapid production of synthetic data, making it a ``semi-amortized'' inference approach \cite{andrew2025amortized}.

\section{Structure functions}
The modeling of the scaling of structure functions are very important in studying turbulent flows. The structure function of
order $n$ is defined as
\begin{equation*}
    \delta_n(h) = E |u(\bs')-u(\bs)|^n,
\end{equation*}
in which $h$ denotes the distance between the points $\bs$ and $\bs'$. In the inertial range, Kolmogorov's law holds and
\begin{equation*}
    \delta_3(h) = M \delta_2(h)^{3/2},
\end{equation*}
where the constant $M$ is independent of $h$; see \cite{stolovitzky1993scaling} for more details. Therefore, examining the structure functions at different levels of $z$ will help us further compare the quality of emulations from POD and xVAE. 

\begin{figure}[!t]
    \centering
    \includegraphics[width=0.9\linewidth]{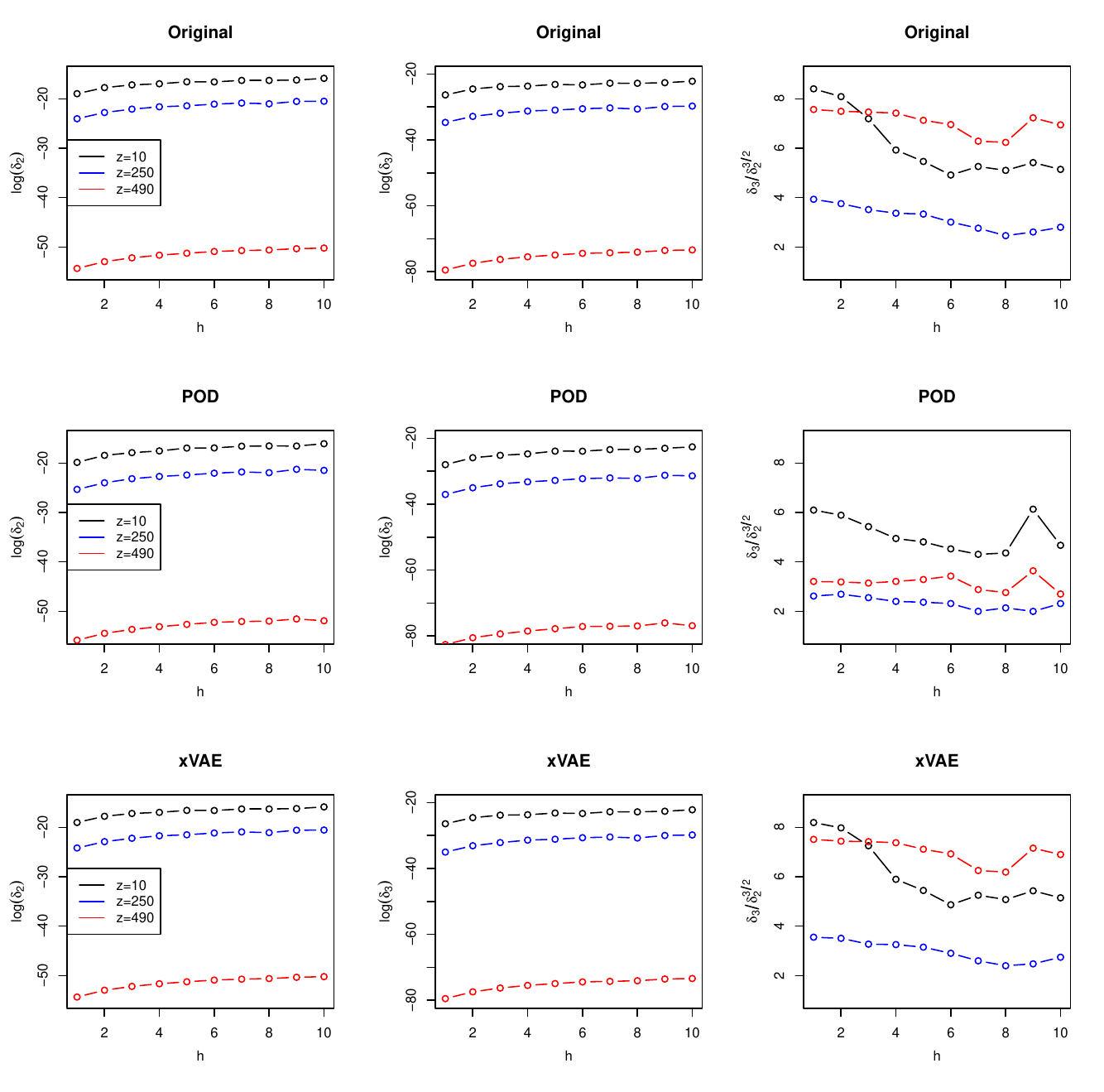}
    \caption{The empirically estimated structure functions of
order $2$ ($\delta_2$; left panels), order $3$ ($\delta_3$; middle panels) and their ratios ($\delta_3/\delta_2^{3/2}$; right panels) from the original data and emulated data from POD and xVAE.}
    \label{fig:struc_fun}
\end{figure}

Figure~\ref{fig:struc_fun} displays the structure functions estimated at levels $z=10$, $250$ and $490$. Both the POD and xVAE emulations produce $\delta_2$ and $\delta_3$ values that closely resemble those from the original plume data.. However, while xVAE maintains the ratio $\delta_3/\delta_2^{3/2}$, consistent with the original data, POD fails to replicate these ratios accurately. This discrepancy indicates that xVAE provides higher-quality emulations. Notably, the POD ratios are consistently lower than the true values, suggesting an underestimation of $\delta_3$ relative to $\delta_2$. Consequently, this implies that the joint occurrence of high/extreme values is underestimated across all length scales (i.e., $h$ from $2$ to $10$).

\section{Additional figures}
\begin{figure}[h]
    \centering
    \includegraphics[width=\linewidth]{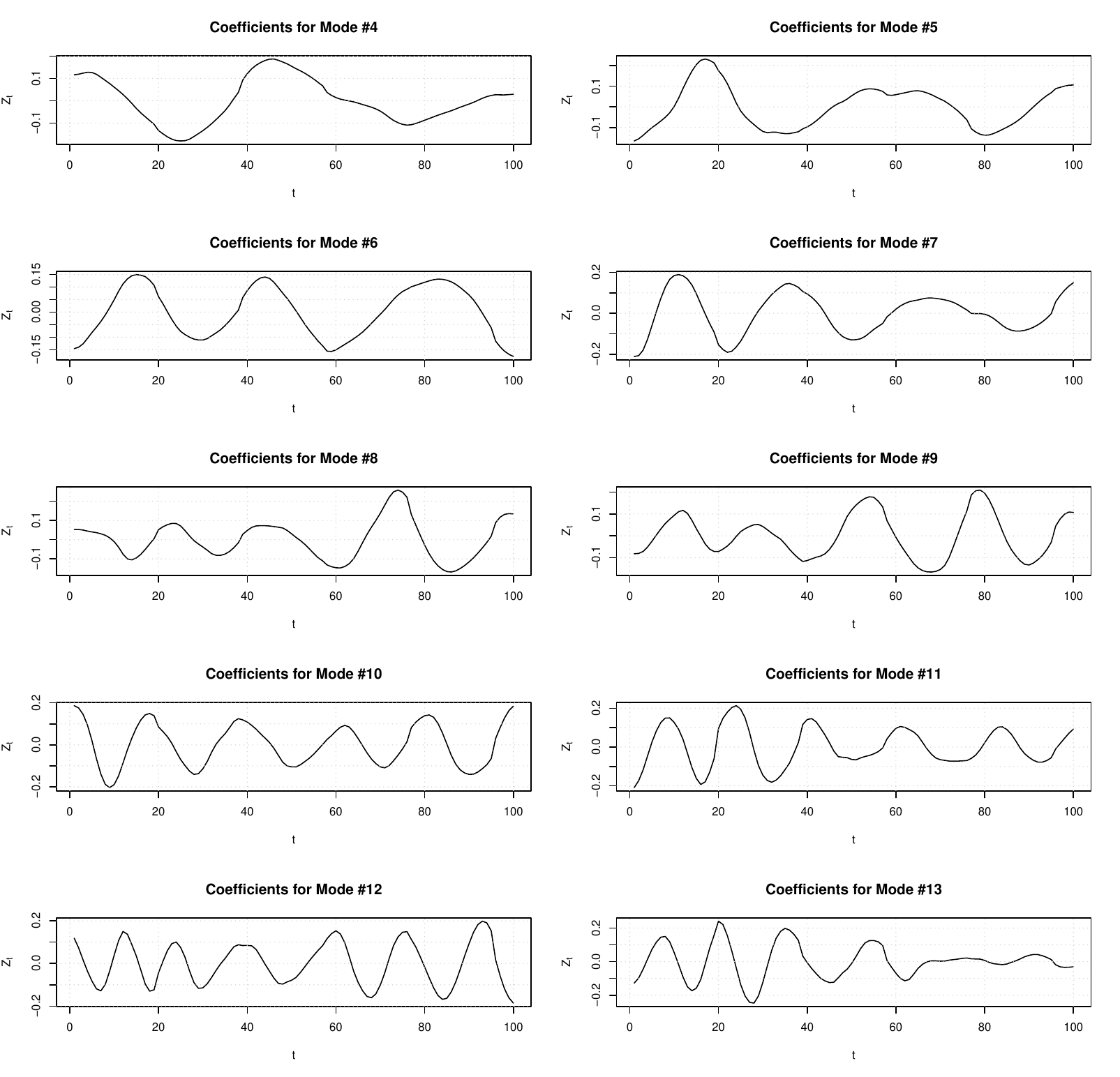}
    \caption{The coefficients $\mathbf{a}_k$, $k=4, \ldots, 10$ from the POD corresponding to the 4--10th modes. See Figure~\ref{fig:pod-features} for the first three modes. We observe that the coefficients exhibit sinusoidal patterns over time  with increasing frequencies as the energy content decreases.}
    \label{fig:pod_coefs}
\end{figure}

\begin{figure}[!t]
    \centering
    \includegraphics[height=0.35\linewidth]{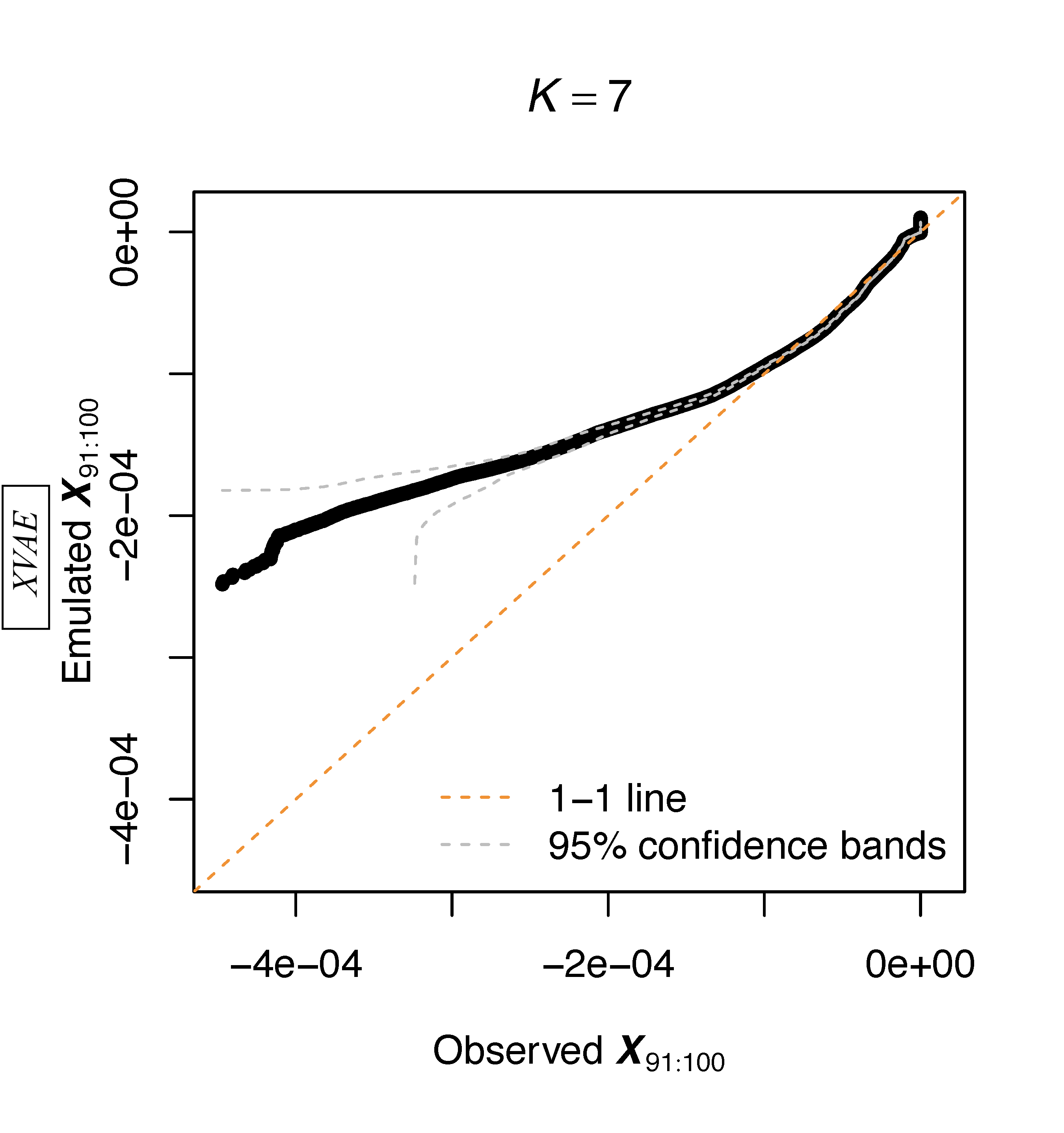}
    \includegraphics[height=0.35\linewidth, trim={1cm 0 0 0}, clip]{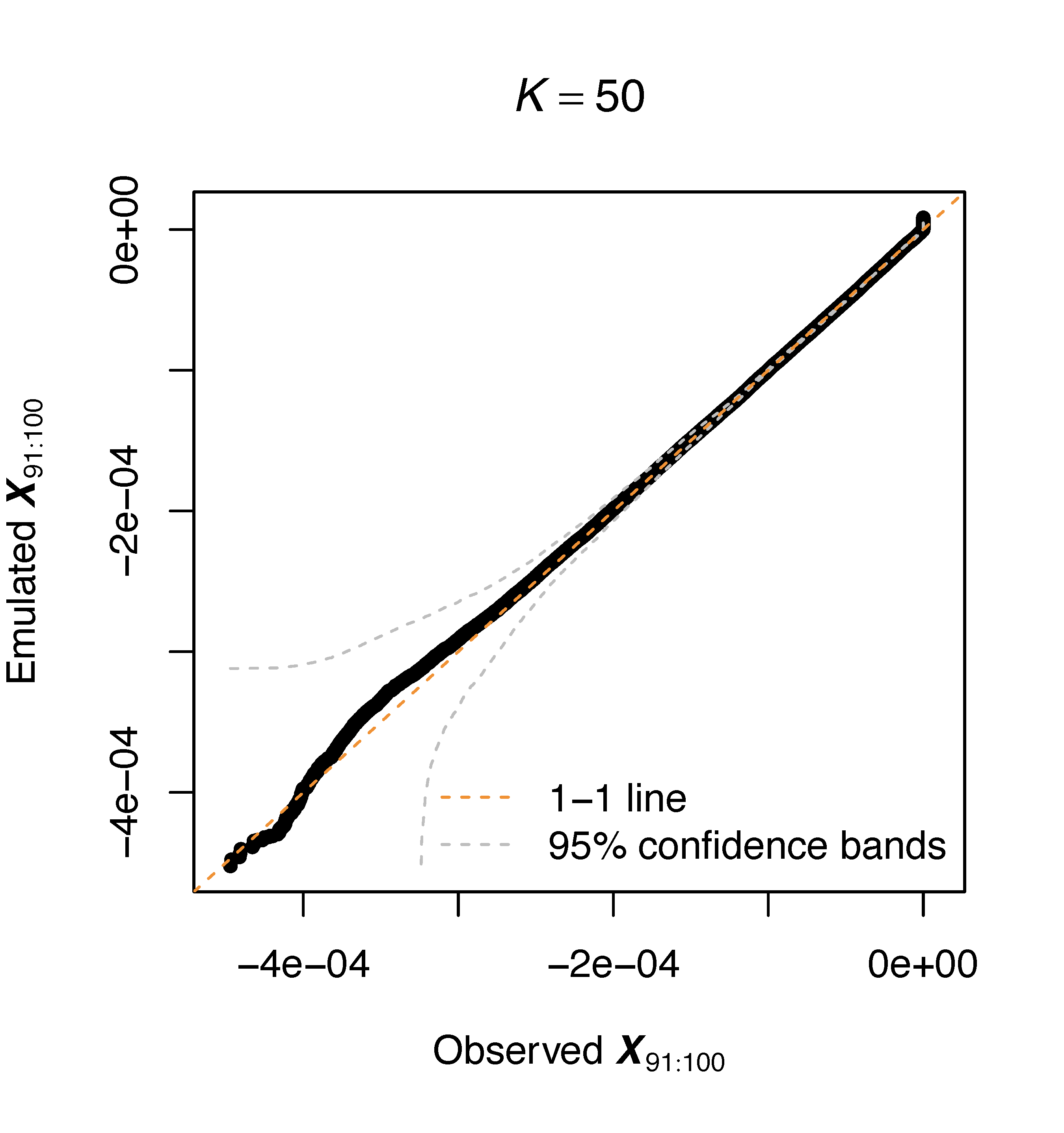}
    \includegraphics[height=0.35\linewidth, trim={1cm 0 0 0}, clip]{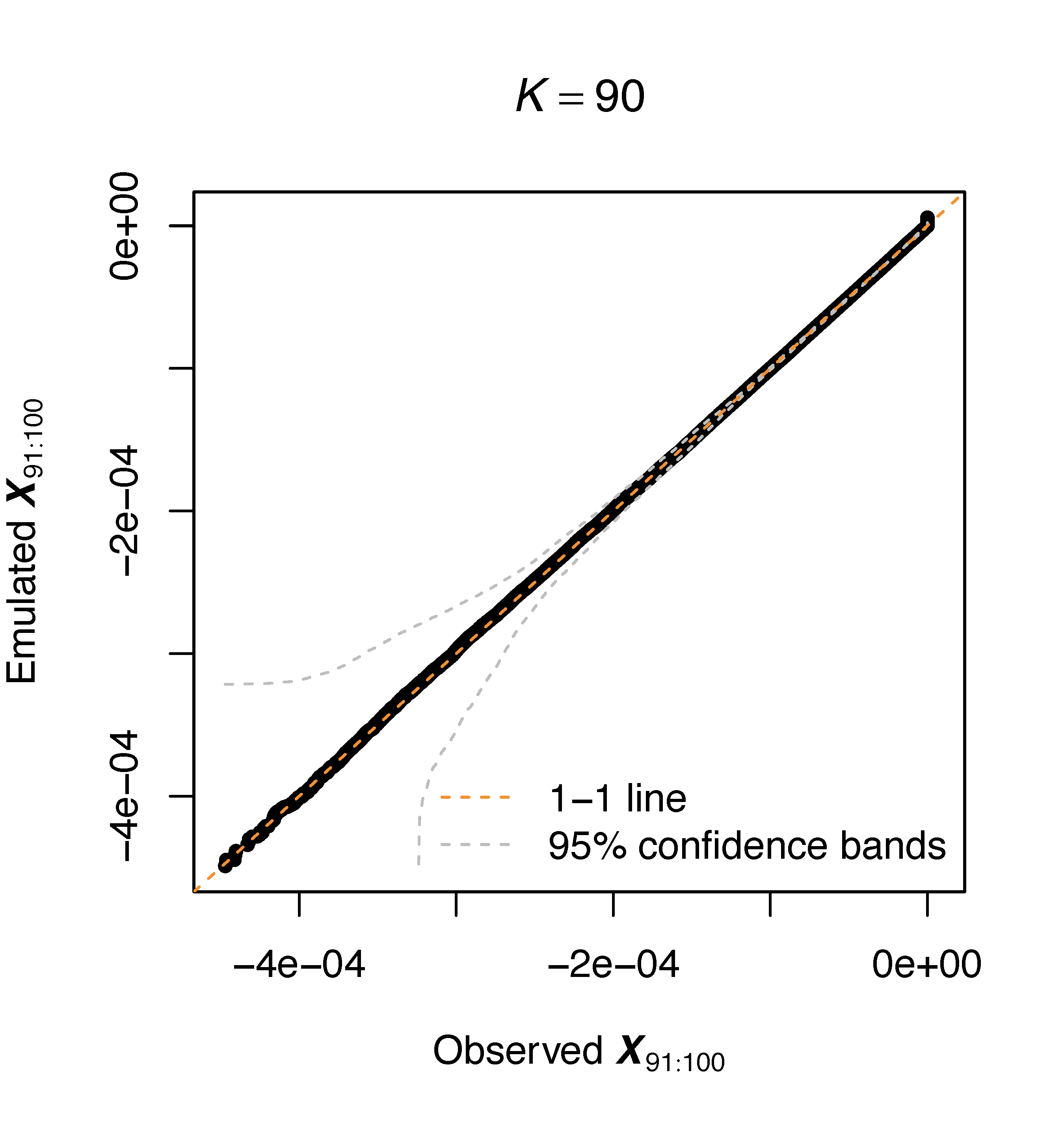}

    \includegraphics[height=0.35\linewidth]{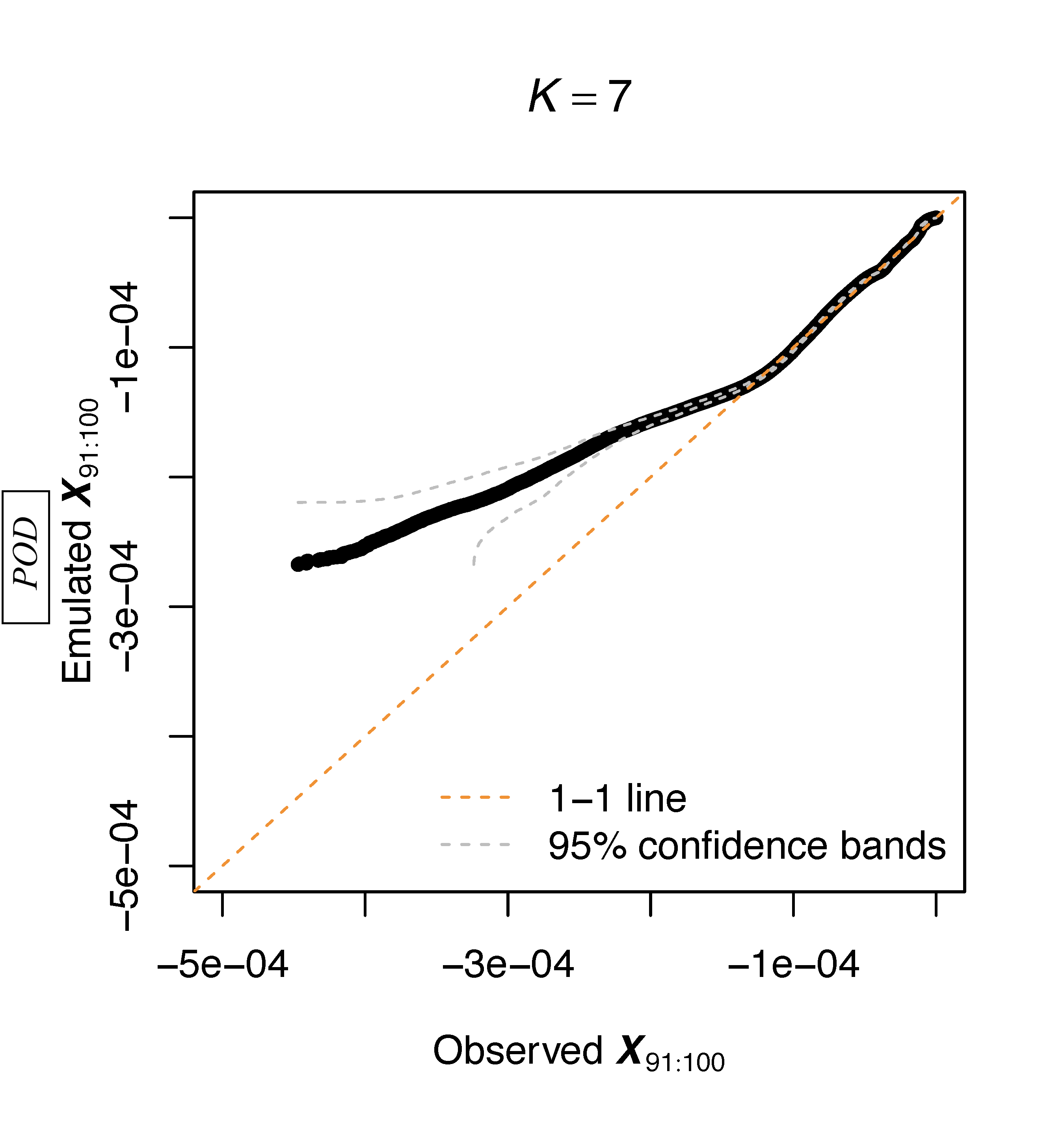}
    \includegraphics[height=0.35\linewidth, trim={1cm 0 0 0}, clip]{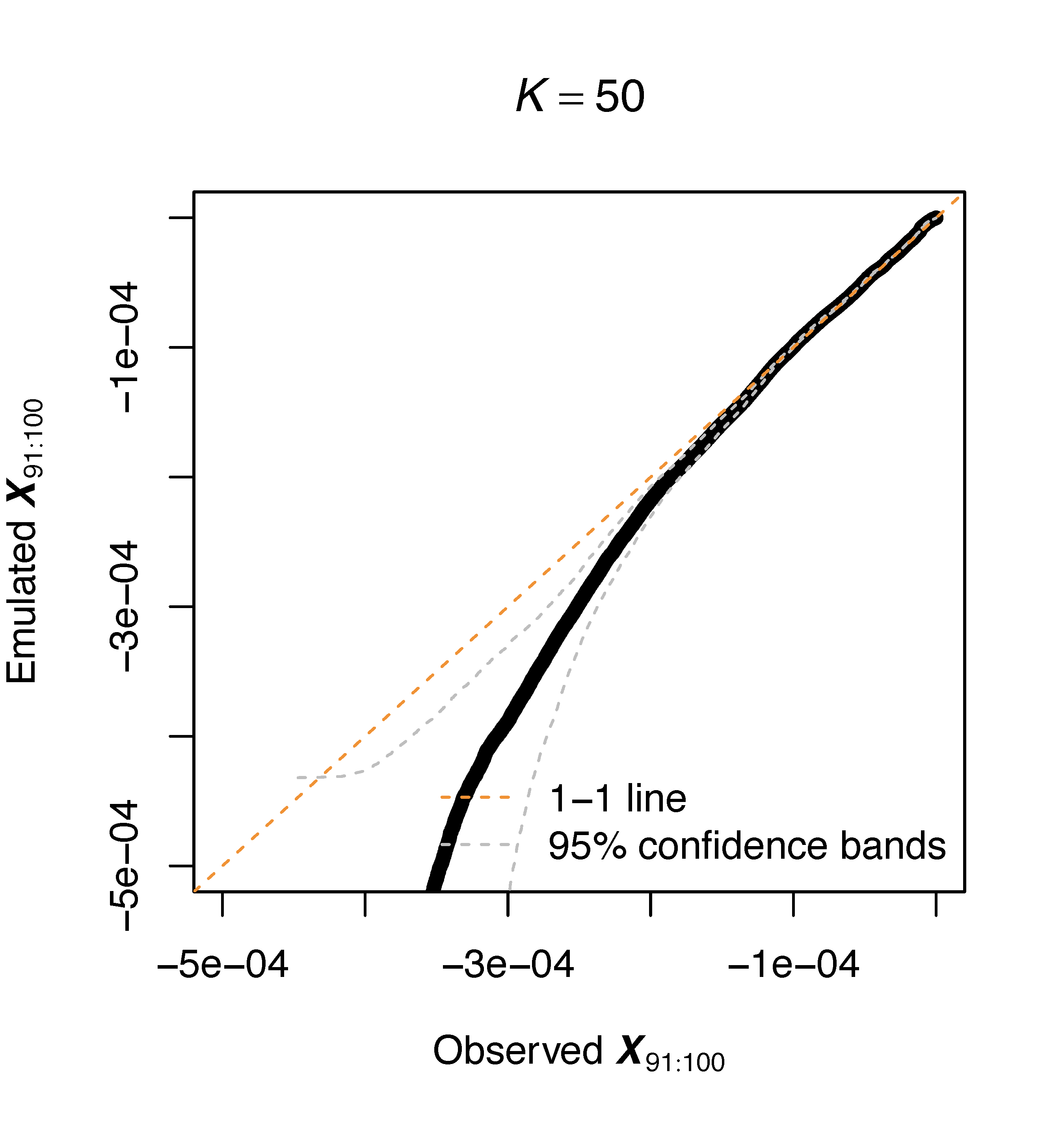}
    \includegraphics[height=0.35\linewidth, trim={1cm 0 0 0}, clip]{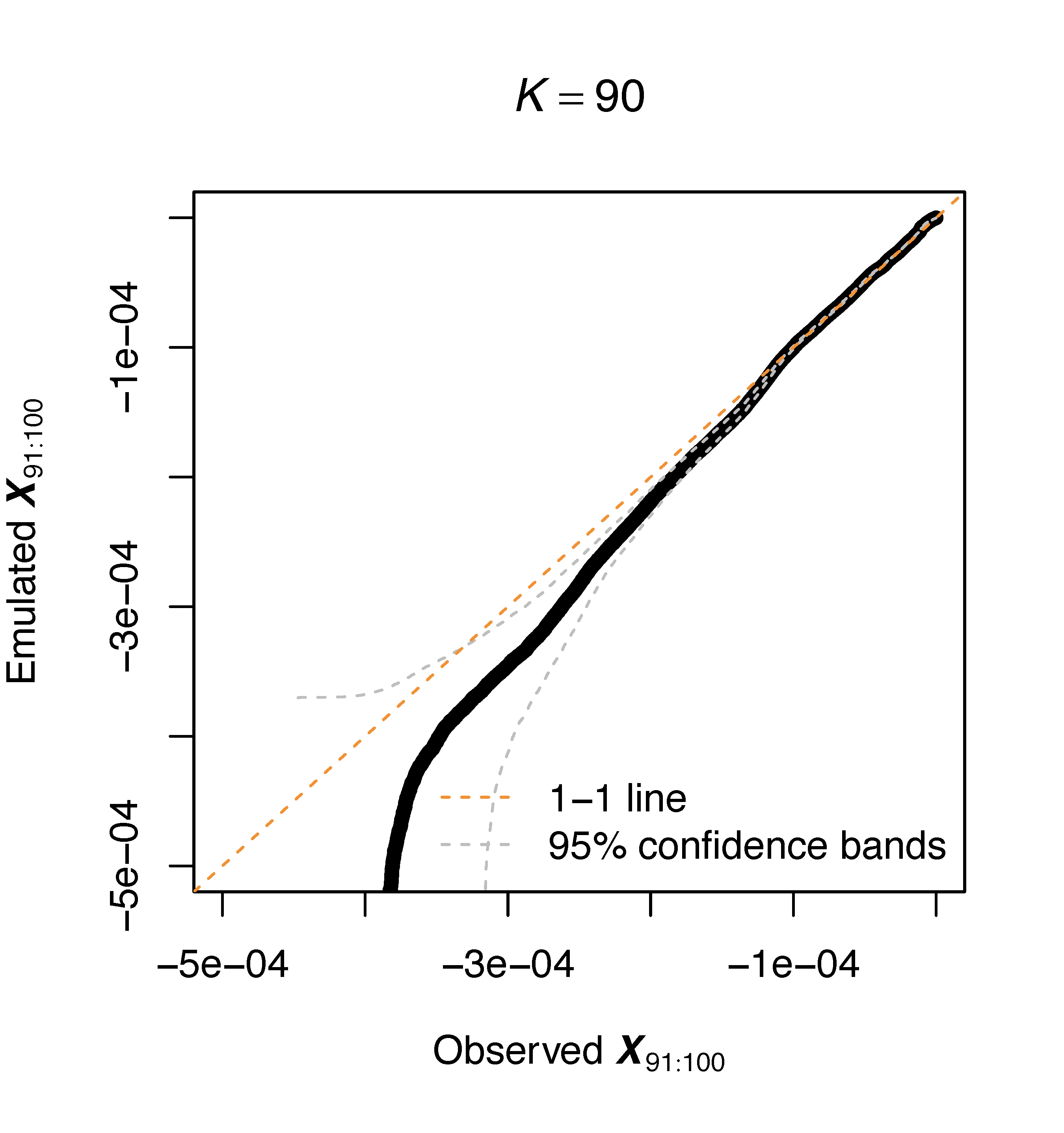}
    \caption{QQ-plots comparing the held-out observed data from fold 10 (i.e., times 91 to 100 are excluded) with the corresponding emulated fields using XVAE (top panels) and POD (bottom panels). The plots illustrate various numbers of bases: $K=7$ (left), $K=50$ (middle), and $K=90$ (right). Here, the 0.001 overall quantile level of $\{\bX_t: t=1,\ldots,10\}$ is approximately -0.00035.  For similar plots corresponding to fold 1, where times 1 to 10 are held out, see Figure \ref{fig:qqplots1} in the main text.}
    \label{fig:qqplots2}
\end{figure}

\begin{figure}[!t]
    \centering
    \includegraphics[height=0.43\linewidth]{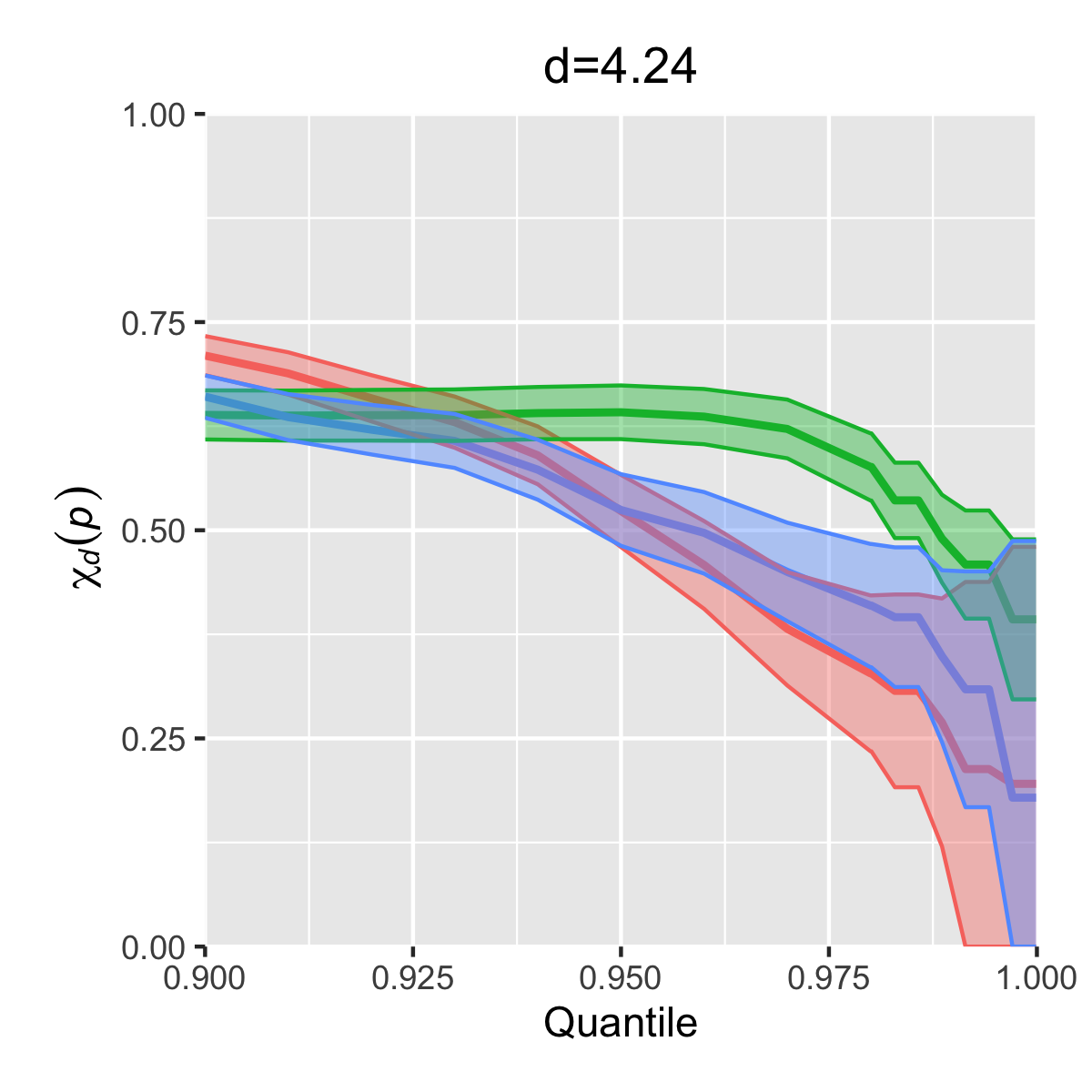}
    \includegraphics[height=0.43\linewidth]{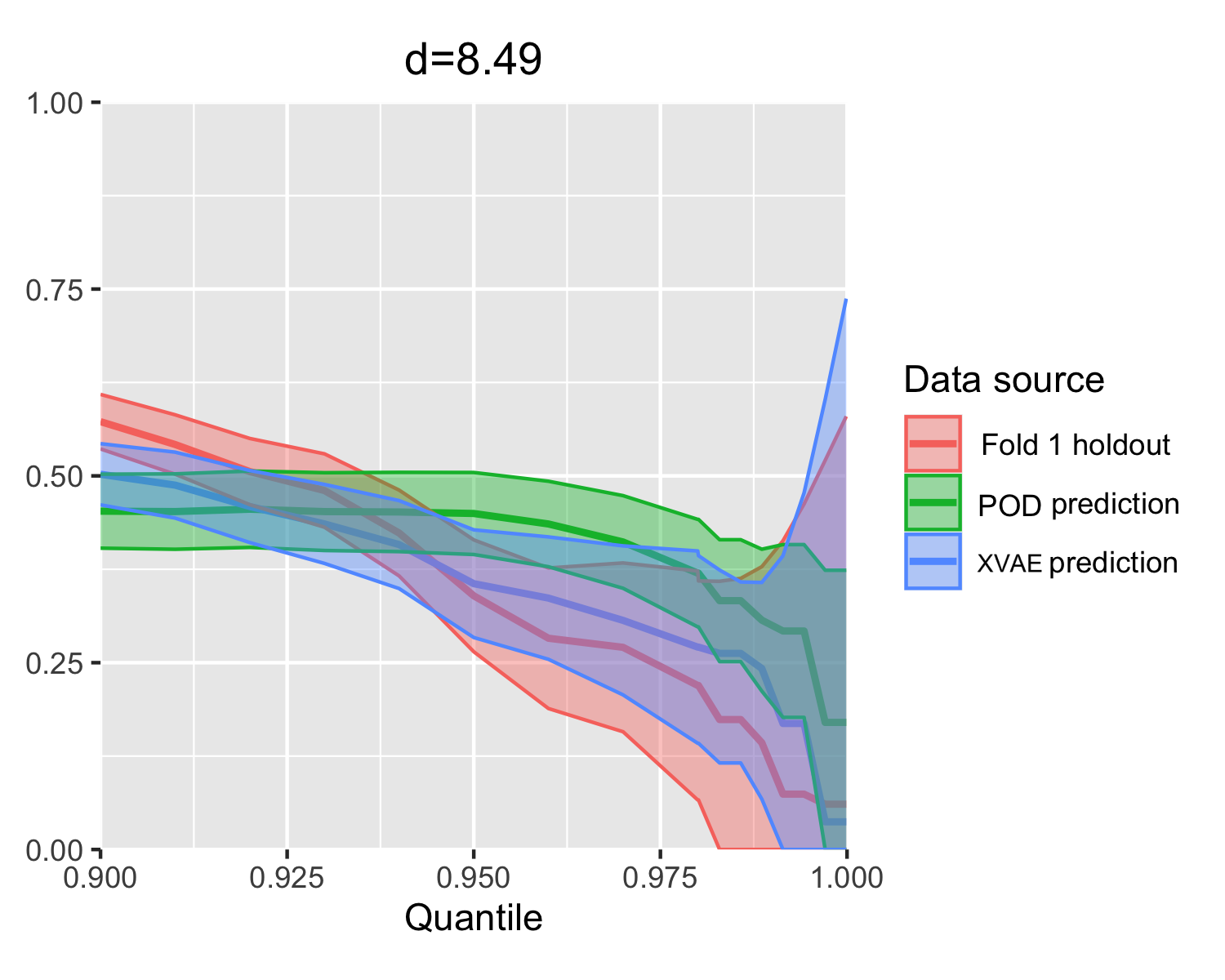}
    \caption{We show the empirically-estimated $\chi_d(p)$ at $d = 4.24$ (left) and $d=8.49$ (right), based on the held-out observed data from fold 1 (i.e., times 1 to 10 excluded), shown in red. The corresponding emulated fields using XVAE are shown in blue, and those using POD are shown in green. For similar results from emulations without held-out data, refer to Figure \ref{fig:chi_plots}.}
    \label{fig:chi_plots_cv}
\end{figure}

\begin{figure}[!t]
    \centering
    \includegraphics[height=0.43\linewidth]{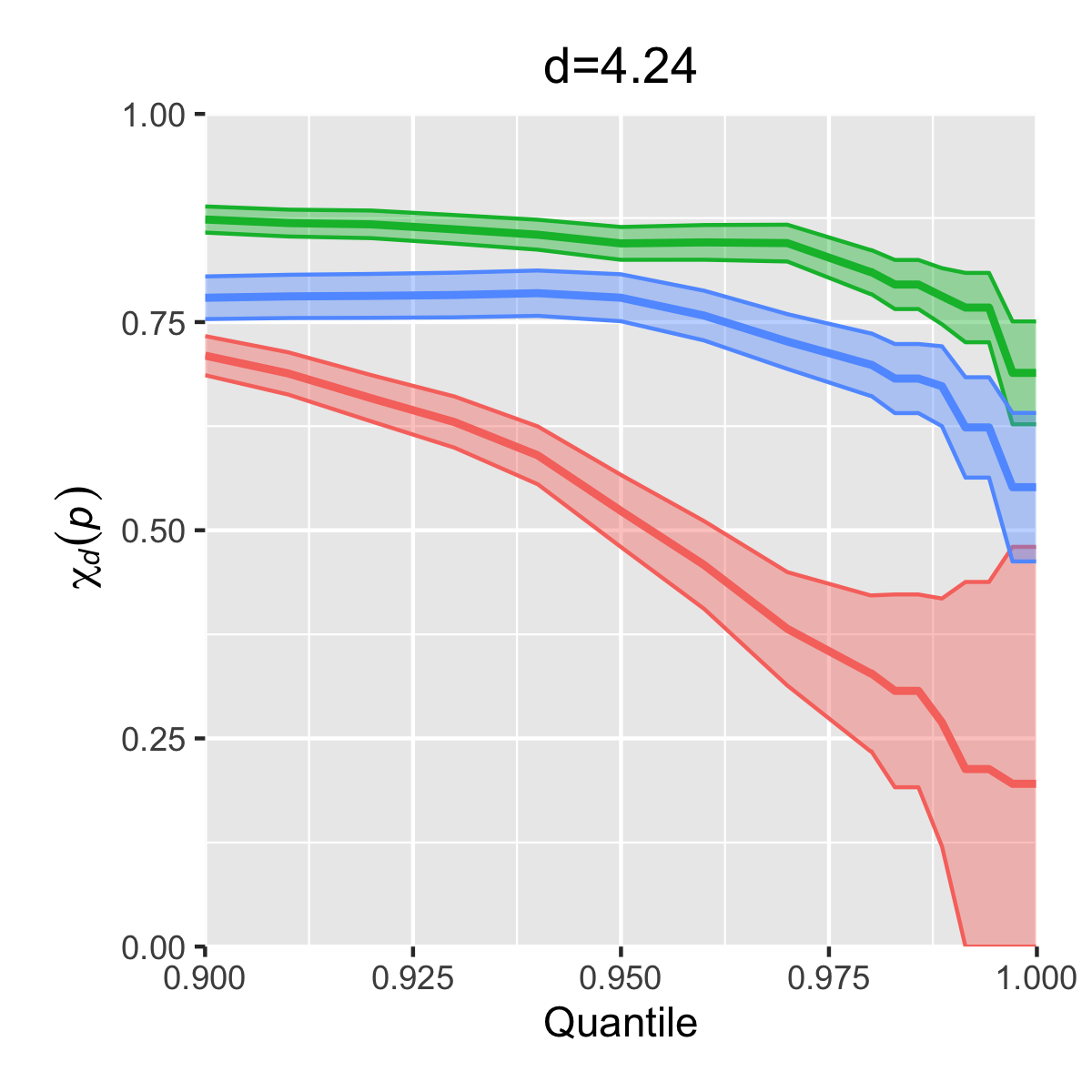}
    \includegraphics[height=0.43\linewidth]{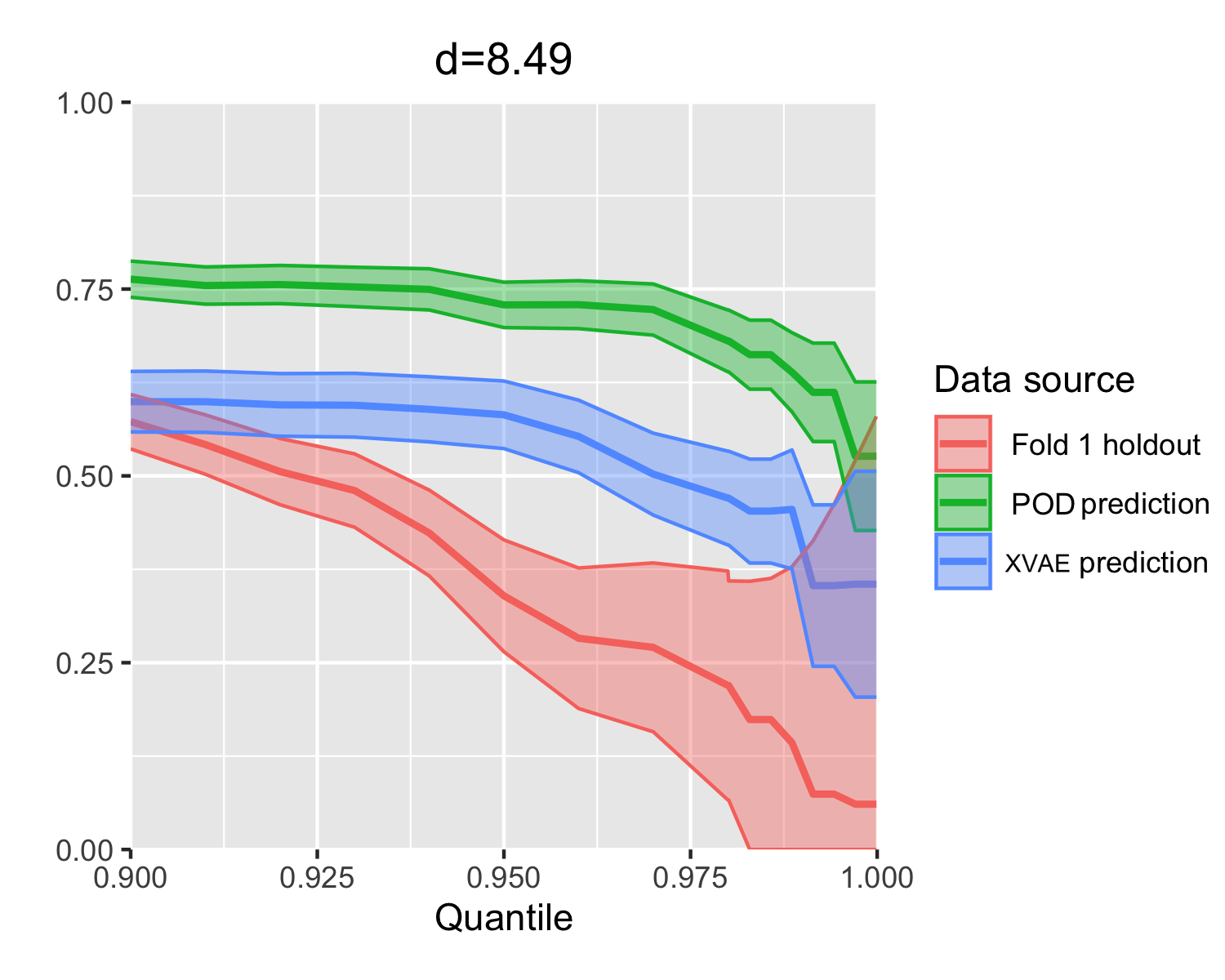}
    \caption{Using 7 principal components, we show the empirically-estimated $\chi_d(p)$ at $d = 4.24$ (left) and $d=8.49$ (right), based on the held-out observed data from fold 1 (i.e., times 1 to 10 excluded), shown in red. The corresponding emulated fields using XVAE are shown in blue, and those using POD are shown in green. For similar results from emulations without held-out data, refer to Figure \ref{fig:chi_plots_ncomp7}.}
    \label{fig:chi_plots_cv_holdout}
\end{figure}
\end{document}